\documentclass[useAMS,usenatbib]{mn2e}
\pdfoutput=1

\usepackage{natbib}
\usepackage{amssymb}
\usepackage{bm}
\usepackage{graphicx}
\let\ni=\noindent

\title[Galaxy clustering using photometric redshifts]{Galaxy clustering using photometric redshifts}

\author[A. M. So\l tan and M. J. Chodorowski]{A. M. So\l tan$^{1}$\thanks{E-mail:
soltan@camk.edu.pl (AMS); michal@camk.edu.pl (MJC)} and M. J. Chodorowski$^{1}$\\
$^{1}$Nicolaus Copernicus Astronomical Centre, Bartycka 18, 00-716 Warsaw, Poland }

\begin{document} 
\date{Accepted 2015 July 21. Received  2015 June 16; in original form 2014 November 20}

\pagerange{\pageref{firstpage}--\pageref{lastpage}} \pubyear{2014}

\maketitle

\label{firstpage}

\begin{abstract}
We investigate the evolution of the galaxy two point correlation
function (CF) over a wide redshift range, $0.2 < z \lesssim 3$.
For the first time the systematic analysis covers the redshifts above
$1 - 1.5$.
The catalogue of $\sim 250000$ galaxies with $i^+ < 25$ and known
photometric redshifts in the Subaru Deep field is used. The galaxies
are divided into three luminosity classes and several
distance/redshift bins. First, the 2D CF is
determined for each luminosity class and distance bin. Calculations
are based on the quantitative differences between the surface
distributions of galaxy pairs with comparable and distinctly
different photometric redshifts.  The power law approximation for
the CF is used. A limited accuracy of photometric
redshifts as compared to the spectroscopic ones has been examined
and taken into account. Then, the 3D functions for all the selected
luminosities and distance are calculated.
The power law parameters of the CF, the slope and
the correlation length, are determined. Both parameters do not show
strong variations over the whole investigated redshift range. The
slope of the luminous galaxies appears to be consistently
steeper than that for the fainter ones. The linear bias factor,
$b(z)$, grows systematically with redshift; assuming the local
normalization $b(0) \approx 1.1-1.2$, the bias reaches $3 - 3.5$
at the high redshift limit.
\end{abstract}


\section{Introduction}

Growth of galaxy structures provides an essential information on the
evolution of the dark matter distribution (\citealt{marulli13}, and
references therein).  Observations of the large scale structures would --
possibly -- give insight also into the very nature of dark energy
(e.g.\citealt{jenkins98,jennings11,huterer15}). Gravitationally dominating
dark matter induces growth of the density fluctuations that eventually lead
to the formation of galaxies. From that moment on, the large scale matter
distribution generated in the computer simulations becomes, at least
potentially, subject to the observational constraints.

However, distinctly different physical properties of the collisionless
dark matter and the visible, baryonic matter make the interplay between
those constituents intricate. Flows of baryonic matter towards
gravitational wells created by concentrations of the dark matter involve
complex processes of gas accretion, shock heating, and radiative
cooling. In effect, the observable galaxy structures do not follow exactly
the dark matter distribution. The relationship between the density
fluctuations of the galaxies and dark matter was formulated on
statistical grounds by \citet{kaiser84}. The amplitude ratio of
these fluctuations, known as the galaxy bias, is roughly linear, although
it depends on the smoothing scale (\citealp{mo96}). The bias is also a
function of time (e.g. \citealp{fry96}). Moreover, the galaxy
clustering depends on galaxy luminosity, mass, colour, and other
parameters. In the local Universe these relationships have been
extensively investigated (e.g. \citealt{coupon12}, see also
\citealt{marulli13}, for the comprehensive reference list).

Complex dark matter structures generated by the gravitational instability
revealed in the cosmological simulations combined with the obscure nature
of the dark matter itself open a space for models that describe the
distribution of luminous matter with the dark one (e.g. \citealt{berlind02,
papageorgiou12}). The study of the luminous -- dark matter relationship
incorporates two separate issues: adequate statistical description of the
galaxy distribution and comprehensive cosmological simulations of dark
matter. Both
determined over a wide redshift range. In the present paper we concentrate
on the galaxy distribution. This question has been investigated extensively
for a long time and in recent years gained momentum mostly as a result of
massive automated galaxy surveys. Although many characteristics of the
galaxy clustering are precisely measured at selected magnitudes and/or
redshifts, the evolution of the clustering over a wide redshift range is
still not well determined. In the present paper we pursue this
question, i.e. to what extent the observational data constrain parameters
of the  cosmic evolution of the galaxy clustering over the whole
observable cosmic time.

In most cases questions related to the galaxy clustering are adequately
addressed using the correlation functions (CFs). Usually the observed
function is satisfactory represented by a power-law with two fitted
parameters -- the correlation length and slope. Only the high-quality
statistical material allows for more detailed study (e.g.
\citealt{baugh96,norberg02, martinez-manso15}). Here we apply the power-law
model for two reasons.  First, the present data do not allow for the more
refined analysis, and second, our method works efficiently using this
approximation. Additionally, we believe that a question of the clustering
evolution is still adequately addressed by the search for variations with
time of the power-law parameters. A large number of investigations in the
recent years examined how the shape of the galaxy CF
depends on the overall galactic parameters such as stellar mass,
luminosity, type, colour or star formation rate. In the local universe and
at low to moderate redshifts, all these studies indicate that the amplitude
of the CF increases with luminosity (e.g.
\citealt{norberg02, pollo06, coil06,li06, wake11, zehavi11, marulli13}),
although strength of this dependence is disputable (\citealt{meneux09}), or
limited to specific types of galaxies (e.g. \citealt{lin12,mostek13}).  On
the other hand, several studies indicate little or no dependence of the
correlation slope on galaxy parameters (e.g.
\citealt{norberg02,coil06,marulli13}), but see \citet{pollo06}.

If no information on the galaxy radial distance is available, the amplitude
of spatial correlations is derived from the 2D CF.
Efficiency of this approach is limited to relatively shallow galaxy
samples.  This is because in the deep galaxy surveys the correlation signal
is diluted by a large number of random coincidences. Galaxy catalogues with
redshifts offer a natural method to eliminate most of the random
coincidences, what substantially improves the signal to noise (S/N) ratio.
Unfortunately, acquisition of the spectroscopic redshifts is time consuming
and restricted to relatively bright objects.

The photometric redshifts, as compared to the spectroscopic ones, are
substantially less accurate distance indicators. Nevertheless, they provide
at least raw estimate of the galaxy position. Thus, photometric redshifts
could be used to identify and extract evident random pairs, and to increase
in this way the S/N ratio of the correlation amplitude measurement. The
photometric redshifts were successfully used in the past for the
investigations of the angular clustering (\citealt{heinis07, mccracken08,
wake11}), and the spatial clustering (\citealt{arnalte-mur14,
bielby14}).
In the present paper we apply a different technique to obtain the
galaxy CF for a wide range of redshifts, not covered by
previous investigations. We use the $2$ deg$^2$ COSMOS photometric redshift
catalogue by \citet{ilbert09} available through the Web site of
IPAC/IRSA.

According to a standard procedure, to determine the 2D CF
one generates a large set of randomly distributed points. Properly
normalized numbers of galaxy-random point pairs is then used as a reference
distribution of pair separations representing a random galaxy population.
A comparison of the physical galaxy-galaxy pairs with the galaxy-random
point pairs allows one to assess fluctuations of the galaxy distribution
and eventually to calculate the correlation signal. The efficiency of this
method is highly sensitive to the interference with the cosmic signal of
various selection effects related to the data processing. For instance,
even a minute variations in the catalogue depth or the image quality result
in fluctuations of the surface density of objects, that could be easily
misidentified with the galaxy clustering. To minimize these kind of
confusions, the 'random' points should be distributed in such a way as to
mimic all the inhomogeneities of the non-cosmic origin. The performance
of this widely applied procedure depends on how precisely such mock
catalogues are free from all the observational biases.

To reduce the instrumental bias, we apply here a different attitude. We
assess the number of galaxy pairs expected for the random distribution by
means of the galaxy-galaxy pairs with sufficiently different photometric
redshifts, that effectively exclude physical connection. Distribution of
such pairs incorporates most signatures associated with the data bias and
processing, while it is free from the physical clustering signal.

The organization of the paper is as follows. In the next section, we give
a short description of the COSMOS photometric redshift catalogue. In
Section~\ref{sec:2Dacf}, we describe the details of the present method to
calculate the 2D CF, derive the relevant formulae and
present results of these calculations.  Formulae applied to determine
the 3D CF for different luminosities and over a
several redshift bins are given in Sect.~\ref{sec:3Dacf}. Also
statistical properties of the photometric redshift measurements are
described in this section. The evolution of bias in the linear model,
and short comparison of our measurements on the CF
evolution with the selected previous results is presented in
Sect.~\ref{sec:conclusions}.  Strong and weak points of our method are
summarized here.

In the paper we consistently  parameterize the COSMOS catalogue data and
results of the investigation using the comoving distances alongside the
redshift. To convert redshifts to the comoving distances, we use the flat
cosmological model with $H_{\rm o} = 70$\,km\,s$^{-1}$Mpc$^{-1}$,
$\Omega_{\rm m,o} = 0.3$ and $\Omega_{\Lambda, \rm o} = 0.7$.


\section{The data}
 \label{sec:data}

\begin{figure}
   \centering
   \includegraphics[width=1.00\linewidth]{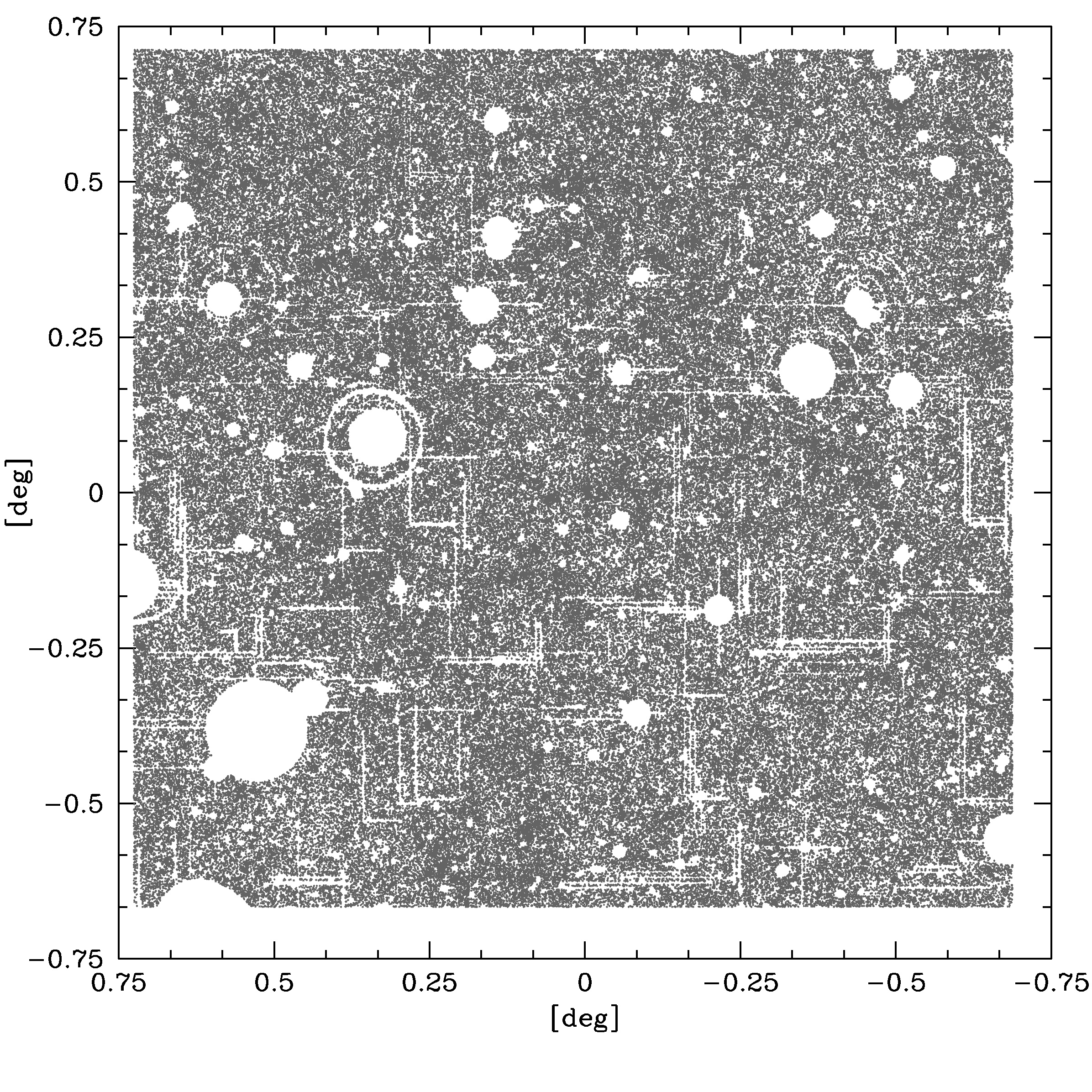}
   \caption{Distribution of $\sim 250000$ COSMOS galaxies. An uneven data
   coverage affects the galaxy correlation analysis.}
   \label{map}
\end{figure}

The COSMOS photometric redshift catalogue is presented in detail by
\citet{ilbert09}. Here, we give only the information relevant to the present
investigation. The catalogue contains $385065$ objects in the deep Subaru
Area, of which almost $252000$ have been classified as galaxies brighter
than $i^+_{\rm AB} = 25$. The galaxies are distributed within a square of
$84$\,arcmin a side centred at $\alpha_c =150\fdg 1$ and $\delta_{\rm c}
= 2\fdg 2$ (\citealt{taniguchi07}). However, the data coverage is
nonuniform on various angular scales. In Fig.~\ref{map}, the distribution
of all the galaxies is shown. One can see a large number of masked areas of
poor image quality (mostly saturated star images and CCD related problems).

To measure the photometric redshifts {\it Le Phare}
\footnote{\scriptsize
www.cfht.hawaii.edu/\textasciitilde{}arnouts/LEPHARE/DOWNLOAD/lephare\_doc.pdf
}
(\citealt{arnouts11})
algorithm was applied using $30$ filters that cover UV, optical and NIR.
According to \citet{ilbert09}, the photo-$z$ accuracy depends on galaxy
redshift and magnitude, and it is suitably characterized by $\sigma_{\Delta
z/(1+z_{\rm s})}$ defined as $1.48 \times {\rm median}(|z_{\rm p} - z_{\rm
s}|/(1+z_{\rm s}))$, where $z_{\rm p}$ and $z_{\rm s}$ denote photometric
and spectroscopic redshift, respectively.  For $i^+_{\rm AB}  < 22.5$ a
dispersion $\sigma_{\Delta z/(1+z_{\rm s})} = 0.007$, while for $i^+_{\rm AB}
< 24$ and $z < 1.25$, $\sigma_{\Delta z/(1+z_{\rm s})} = 0.012$.  At fainter
magnitudes the estimates are less accurate, e.g. for $i^+_{\rm AB}  \sim 24$
and $z \sim 2$, $\sigma_{\Delta z/(1+z_{\rm s})} = 0.06$.

Although the distribution of differences $(z_{\rm p} - z_{\rm
s})/(1+z_{\rm s})$ is roughly fitted by a Gaussian function, the median
statistics in the $\sigma_{\Delta z/(1+z_{\rm s})}$ definition above indicates
that gross errors affect occasionally the $z_{\rm p}$ estimates.
\citet{ilbert09}
define a `catastrophic failures' of the $z_{\rm p}$ estimate if $|z_{\rm
p} - z_{\rm s}|/(1+z_{\rm s}) > 0.15$.  In the case of bright galaxies
($i^+_{\rm AB}  < 22.5$), the fraction of catastrophic failures amounts to
$0.7$ percent. It rises however, to $20$ percent for galaxies at $1.5 <
z_{\rm s} < 3$. The median apparent magnitude of those galaxies $i^+_{\rm
AB} \sim 24$.

\begin{figure}
\centering
\includegraphics[width=1.00\linewidth]{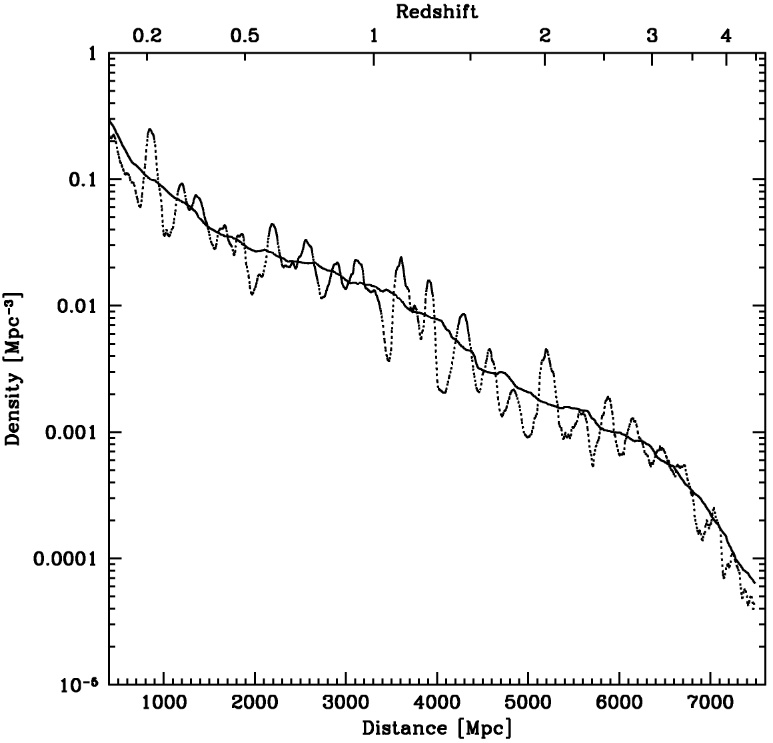}
\caption{The spatial density of the COSMOS galaxies as a function of the
comoving distance (bottom abscissa) or redshift (top abscissa).  Highly
fluctuating (dotted) curve represents the moving average calculated
within $\pm 50$\,Mpc, while the smoother one (solid) -- within $\pm
500$\,Mpc.}
\label{cmv_dens}
\end{figure}

\begin{figure}
   \centering
   \includegraphics[width=1.00\linewidth]{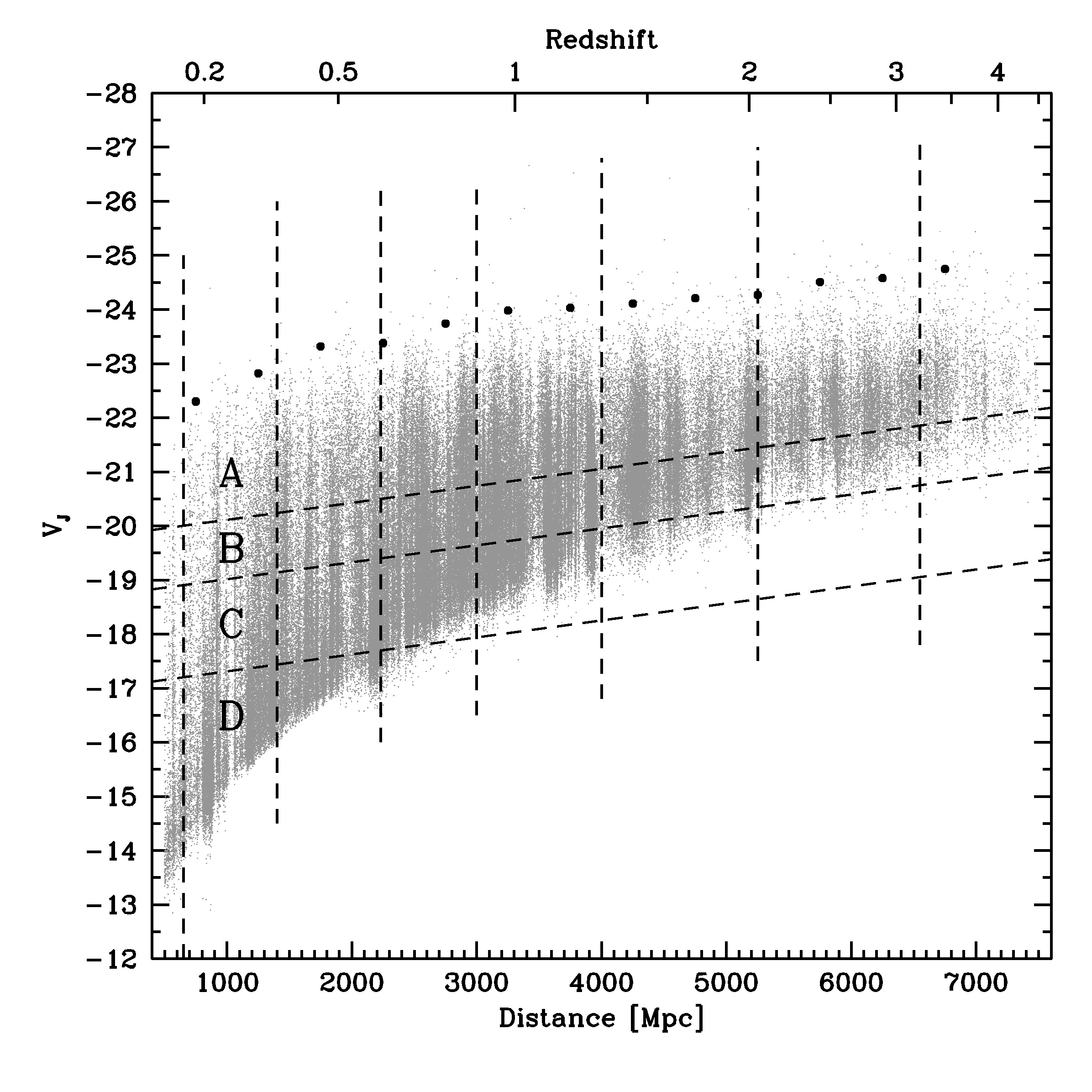}
   \caption{The distribution of galaxies in the absolute $V_J$
            magnitude -- comoving distance plane. Inclined dashed lines
            indicate luminosity sub-samples A--D, while the vertical
            lines define distance bins used in the calculations.
            Full dots - see text.}
   \label{mag_d}
\end{figure}

The distribution of $z_{\rm p}$ over $z_{\rm s}$ exhibits also some
peculiarities that are not depicted by $\sigma_{\Delta z/(1+z_{\rm s})}$.
The $z_{\rm p} - z_{\rm s}$ differences are correlated with $z_{\rm s}$
over scales comparable and larger than the quoted above dispersion
$\sigma_{\Delta z/(1+z_{\rm s})}$. In the right-hand panel of fig.~10 in
\citet{ilbert09} the distribution of $z_{\rm p}$ versus $z_{\rm s}$
exhibits systematic, to some extent coherent variations around the line
$z_{\rm p} = z_{\rm s}$, indicating the non-random, large-scale
deviations between $z_{\rm p}$ and $z_{\rm s}$.  The effect of the
non-random character of $z_{\rm p}$ deviations from the spectroscopic
data is graphically demonstrated in Fig.~\ref{cmv_dens}, where we plot
a moving average density of galaxies in the COSMOS catalogue. We count
galaxies within $50$\,Mpc of the radial distance for each object in the
catalogue. The $z_{\rm p}$ redshifts are used in the calculations.
Number of neighbours is then used to assess the local spatial
concentration of galaxies as a function of distance. Apparent
quasi-periodic oscillations around the average galaxy density in the
distance range $\sim\!2000$ through $\sim\!7000$\,Mpc with the
characteristic length of $\sim\!330$\,Mpc demonstrate the large-scale
inhomogeneities in the $z_{\rm p}$ measurements.

Both the $z_{\rm p}$ inaccuracies and the sky coverage discontinuities
introduce multiple biases in the COSMOS galaxy data that distort the
actual spatial structures. Although it is feasible to construct the
mask that would eliminate (all\,?) the area not covered by the COSMOS
catalogue, the remaining surface nonuniformities potentially present
in the data would persist. In the next section we present a practical
method how to isolate the real clustering signal from all the
non-cosmic effects.

The sample spans a wide range of absolute magnitudes. In
Fig.~\ref{mag_d} the absolute $V_J$ magnitudes (\citealt{ilbert09}) are
plotted against the distance. Apart from the clear effects introduced
by the $z_{\rm p}$ distortions, an inclination of the high-luminosity
envelope defines a rate of the average galaxy luminosity evolution.
Black dots in Fig.~\ref{mag_d} show the magnitudes of $10$\,th ranked
galaxy in $500$\,Mpc bins between $500$ and $7000$\,Mpc. Although in
the course of the cosmic evolution galaxy luminosities are subject to
more complex variations, we adopt here this relationship as the
luminosity evolution of the general galaxy population. We divide the
data into absolute magnitude classes using lines of fixed slope roughly
consistent with the slope of the $V_J(10)$ -- distance relationship
(see Fig.~\ref{mag_d}). The dashed lines in the figure define four
luminosity sub-samples.  The samples A--C contain $\sim 70000$
galaxies each, while the sample D -- around half of this amount.
Because the  D sample is limited to redshifts smaller than $0.5$,
the present analysis of the clustering evolution concentrates on
samples A--C.
Although the rate of the luminosity evolution adopted here is linear in
the comoving distance, it is in good agreement with the evolution model
(linear in redshift) adopted by \citet{marulli13} in the redshift range
of $0.5$--$1.1$.

Moving to larger distances all the galaxy sub-samples become increasingly
incomplete (Fig.~\ref{cmv_dens} for the whole data, and Fig~\ref{bin_dens}
for the class A). This effect should not coerce the measurements of the
CF if the magnitude selection at a given distance
is uncorrelated with the local galaxy space density, and this is assumed
in the subsequent calculations. 


\section{2D correlation functions}
\label{sec:2Dacf}

We take the photometric redshifts, $z_{\rm p}$, as a working estimate of
the comoving distances for all the galaxies. This is a legitimate
assumption for the majority of galaxies. According to \citet{ilbert09},
only a (small) fraction of $z_{\rm p}$ differs from $z_{\rm s}$ by more
than $0.15\cdot (1 + z_{\rm s})$. Nevertheless, in the present derivation
we explicitly take into account a questionable nature of the individual
distance estimate. This is because even a modest fraction of catastrophic
errors affects the present analysis (see below).

Clearly, the distance estimates based on $z_{\rm p}$ are too
coarse to measure directly the spatial clustering. Nevertheless, the
$z_{\rm p}$ data are valuable in measurements of the galaxy clustering at
different distances using the 2D CF. Let us to consider
the expected distribution of galaxies around the galaxy selected at a
distance $d = d(z_{\rm p})$.
The expected number of galaxies within a solid angle $\Delta \omega$
at the angular distance $\theta$ from the selected galaxy, $N(\theta)$,
is described by the 2D CF $w(\theta\, |\, d)$:

\begin{equation}
N(\theta) = \Delta \omega\, [n_{\rm o}\cdot w(\theta\, |\, d) + n_{\rm o}]\,,
\label{first}
\end{equation}

\ni where $n_{\rm o}$ is the galaxy surface density expected in the absence
of cosmic clustering. For the perfect data $n_{\rm o}$ is equal to the mean
galaxy surface density. However, the present observational material reveals
numerous defects that interfere with the sky fluctuations. In effect, the
local `background' galaxy density, $n_{\rm o}$, is not constant but
depends on the position of selected galaxy and the separation $\theta$.

For the data spanning a large distance range as in the case of the COSMOS
catalogue, equation~(\ref{first}) is of a limited use  because the excess of
neighbours, $\delta N(\theta) = N(\theta) - \Delta \omega\cdot n_{\rm o}$,
even at small separations $\theta$, is tiny as compared to
$\Delta\omega\cdot n_{\rm o}$. To improve the S/N ratio, which is of the
order of $\delta N(\theta)/\!\sqrt{N(\theta)}$, we divide the whole galaxy
sample into into six distance bins between $650$ and $6550$\,Mpc.
The radial depth of each bin is larger than $750$\,Mpc. Thus, it is also
much larger than the maximum distance at which the CF differs from $0$.

Calculating the surface correlations of galaxies within a selected bin, we
split the whole galaxy population into two classes. Class I contains
galaxies located in the bin of the selected galaxy, while the class II
contains all the galaxies in other bins. The expected average excess of
galaxies, $N_{\rm I}(\theta)$  or the surface density profile, $n_{\rm
I}(\theta)$, of class I galaxies in the vicinity of the galaxy drawn from
the same distance bin is described by the CF $w_{\rm I}(\theta)$:

\begin{equation}
\frac{N_{\rm I}(\theta)}{\Delta \omega} = n_{\rm I}(\theta) =
n_{\rm Io} \cdot w_{\rm I}(\theta) + n_{\rm Io}\,,
\label{wI}
\end{equation}

\ni where $n_{\rm Io}$ is the class I galaxy surface density expected for
the non-clustered case. It is subject to various observational
constraints and it may vary alike $n_{\rm o}$.  We now assume a power law
for the $w_{\rm I}(\theta)$ function separately for each bin:

 \begin{equation}
w_{\rm I}(\theta) = w_i\, \theta^{\,\zeta_i}, 
                                     \hspace{10mm} i=1,\:.\: .\: ., 6\,.
\label{power_law}
\end{equation}

\ni In the absence of all the effects involved in the data acquisition
that hamper the genuine sky distribution of galaxies, one could use
directly equations~(\ref{wI}) and (\ref{power_law}) to determine the amplitude,
$w_i$, and slope, $\zeta_{\rm i}$. However, in the real data the
`average' galaxy density, $n_{\rm Io}$, is not well defined.
One way to eliminate effects of the $n_{\rm Io}$ fluctuations is to
use the surface distribution of class II galaxies, $n_{\rm IIo}$.
We assume that intruding (non-cosmic) factors that affect the surface
distribution of class I galaxies modify also distribution of class II
galaxies. Although a response of both galaxy population to various
effects may not be identical, we assume that the ratio
$\eta_i = n_{\rm Io}/n_{\rm IIo}$
is much more immune to observational
biases than $n_{\rm Io}$ and $n_{\rm IIo}$ separately.
Thus, dividing equation~(\ref{wI}) by $n_{\rm IIo}$, we get the numerically
tractable equations:

\begin{equation}
\frac{N_{\rm I}(\theta)}{N_{\rm II}(\theta)} =
\eta_i\, w_i \theta^{\,\zeta_i} + \eta_i\,,
\label{basic}
\end{equation}

\ni where $N_{\rm II}(\theta) = \Delta \omega\cdot n_{\rm IIo}$ is the
average number of class II galaxies at the distance $\theta$ from the
randomly chosen galaxy of class I.  As an estimator of $N_{\rm
I}(\theta)$ and $N_{\rm II}(\theta)$ for the given distance bin we use
the total numbers of the class I galaxy pairs and the class I--class II
pairs, respectively, both summed over the entire field.

To assess the power-law parameters of the angular CF, the present
method applies the pair count ratio $n_{\rm Io}/n_{\rm IIo}$ ratio
rather than just the pair counts $n_{\rm Io}$, This scheme deals
simultaneously with two questions. First, it accounts in a simple way
for all the masked out areas. Second, if there are some unrecognized
systematic effects that perturb the surface galaxy distribution, they
are absorbed by the $\eta_i$ parameter while the pure clustering signal
is represented by the power law.

\begin{figure}
   \centering
   \includegraphics[width=1.00\linewidth]{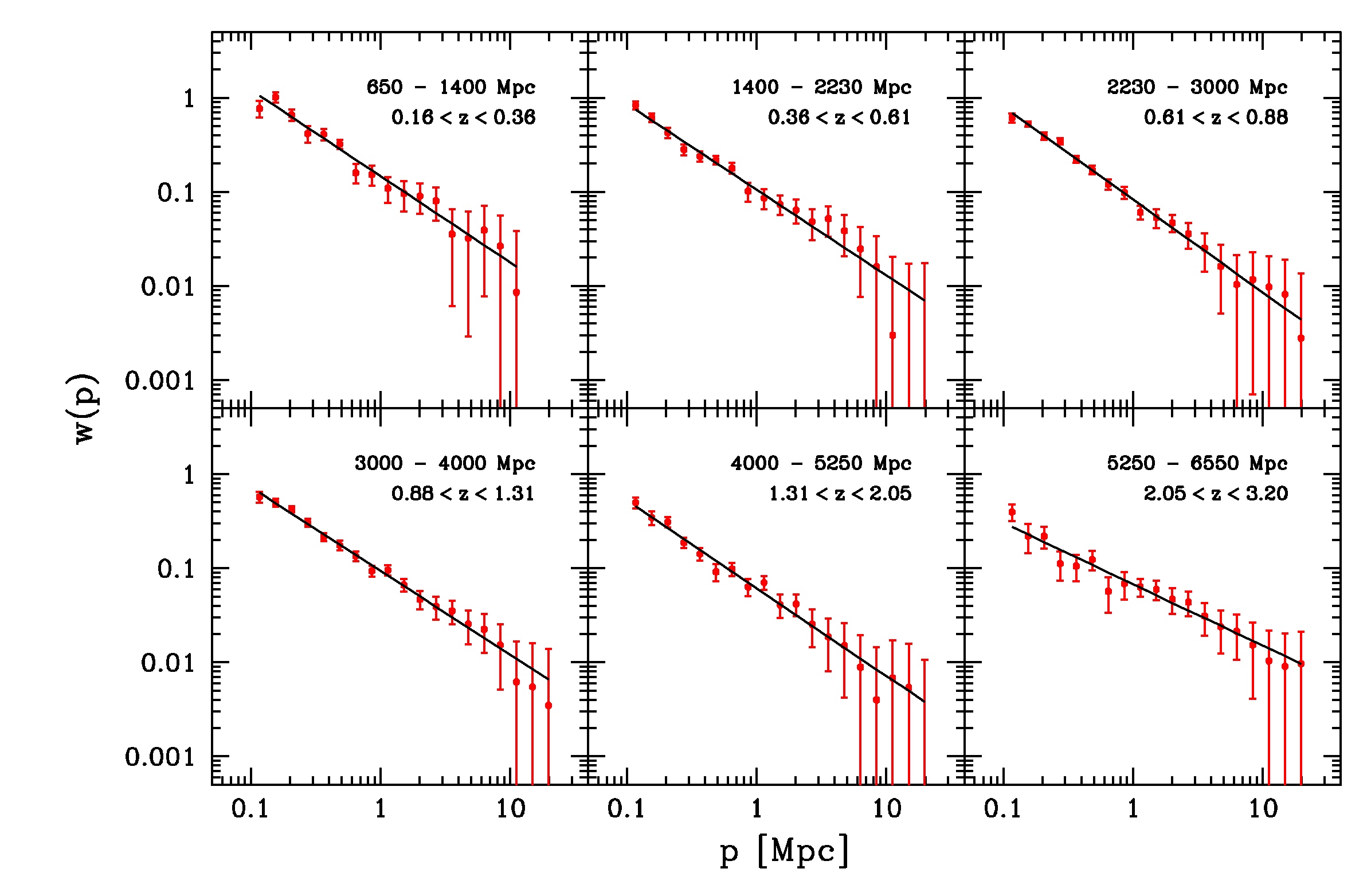}
   \caption{The surface correlation functions for the luminous galaxies
   (sample A) in the six distance bins (labelled in the bottom left
   corner),
   the distance/redshifts boundaries are indicated in the top right
   corner. The data points, $1\sigma$ error bars, and the power-law
   least-squares solutions of equation~(\ref{basic}) are shown.}
   \label{w2_all}
\end{figure}

The parameters $w_i$, $\zeta_i$ and $\eta_i$ for each
distance bin $i$ are obtained as the iterative least-squares solution of
equation~(\ref{basic}). Two first parameters, $w_i$ and $\zeta_i$,
define the 2D CFs. The results for the luminosity
sample A are shown in Fig.~\ref{w2_all}. To ease the comparisons of the
correlation parameters in the different bins, and to indicate what
linear distances are involved in the calculations of the space
CF (see below), we plot the data as a function of the
transverse comoving distance, $p$, rather than the angular separation,
with $p = \theta\cdot R_i$, where $R_i$ denotes the distance
to the centre of the $i$th bin.  In all the distance bins, except for
the nearest one, the galaxy pairs are counted in $19$ separation zones
in the range of $0.1 < p < 22.6$\,Mpc. In the nearest distance bin
($650$-$1400$\,Mpc) the range of separations is reduced to $0.1 < p <
10.0$\,Mpc due to the small angular size of the COSMOS field.

\begin{figure}
   \centering
   \includegraphics[width=1.00\linewidth]{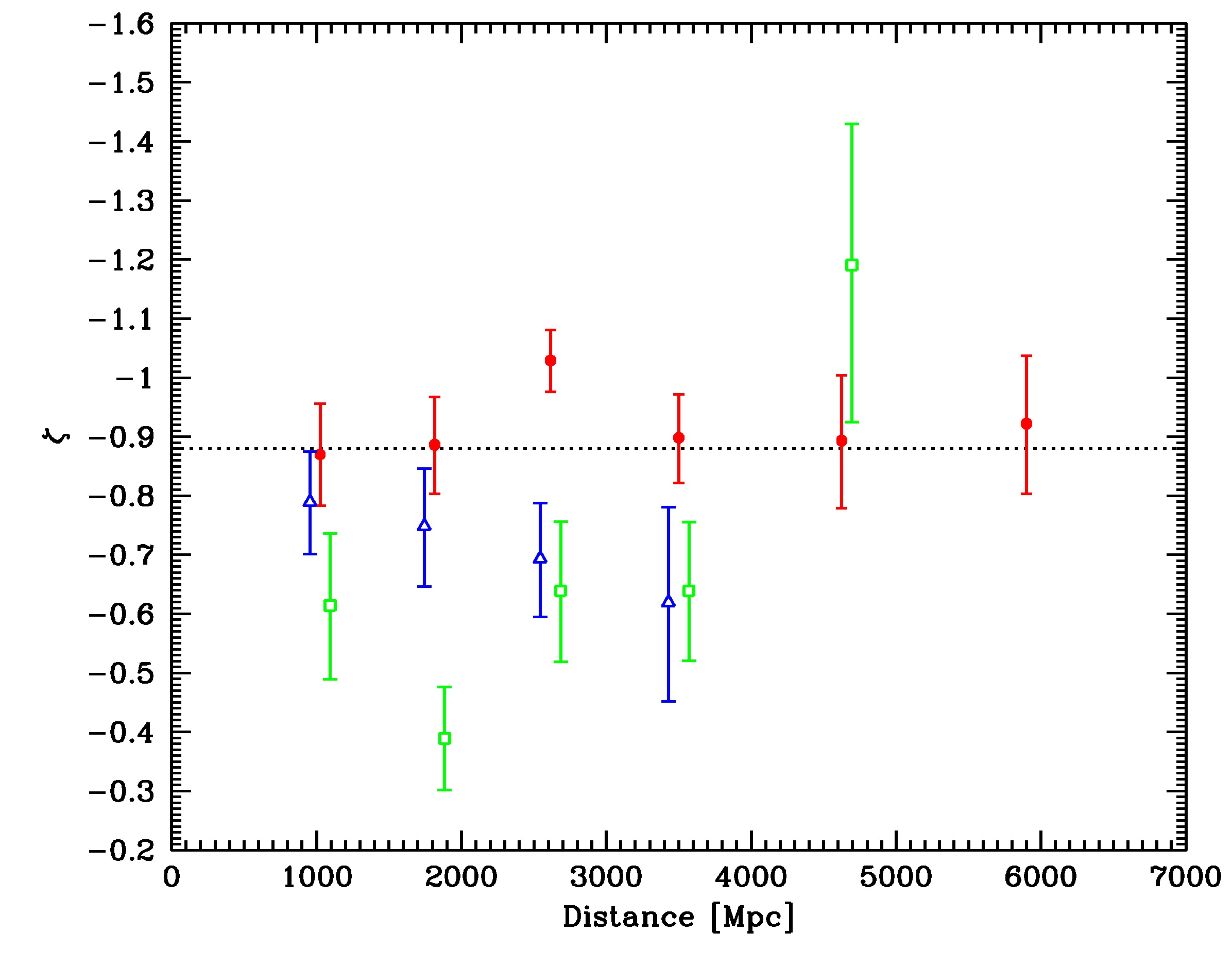}

   \caption{The 2D correlation function slopes: full points -- luminosity
            class A, open circles -- class B, triangles -- class C.}
   \label{w2_slopes}
\end{figure}

The parameters $w_i$ depend on the actual 3D clustering amplitude
and  on the surface density of galaxies in the consecutive distance
bins. To assess the spatial clustering one requires the information on
the spatial density of galaxies in the each bin, and this question is
discussed below. The CF slopes, $\zeta_i$, are
not greatly affected by the bin limits. In Fig.~\ref{w2_slopes}
variations of the correlations slopes as a function of distance are
shown for galaxies in three luminosity samples. The error bars indicate
$68$\,\% confidence levels assuming one interesting parameter
(\citealt{avni76}).  Large $\zeta_{\rm i}$ uncertainties in the
samples B and C make any definite account on the slope variations
problematic. Nevertheless, two conclusions seem to be relatively well
established. First, no obvious trend of slope changes with the distance
is present in any of the galaxy luminosity samples.  Second, the slope
for the most luminous galaxies (sample A) is generally steeper than that
for the remaining ones. The detailed interpretation of the present
results on CF slope is discussed jointly with the amplitude of the spatial
CF in Section~\ref{sec:conclusions}.

   \vspace{5mm}

\section{3D correlation functions}
\label{sec:3Dacf}

We use the  photometric redshifts solely to define wide distance
bins, and not to examine individual galaxy pairs in 3D. The excess
number of close pairs relative to that expected for the random
distribution is clearly visible in each bin. The fine 2D correlation
signal shows that the statistical characteristics of photometric
redshifts are sufficient to determine the evolution of the CF slope over
a wide range of cosmological epochs. To assess the amplitude of the
space correlations at successive distances, one should deproject the
corresponding surface CFs.  

The space CF gives the number of excess galaxy pairs relative to
the local average galaxy density. In the present investigation the
galaxy density varies between bins and also within each individual bin.
However, these variations do not constrain the CF estimates. The
galaxy excess given by the 2D CF is equal to the integral of the space
3D properly weighted by the space density of galaxies populating the
selected distance bin. Below the deprojection procedure is described in
detail.

A power law provides not only satisfactory approximation to the 2D CF,
but also allows for a straightforward assessment
of the spatial correlation parameters. Under the standard assumption of
clustering isotropy and small angles approximation, the spatial
CF is a power law with a slope $\gamma = \zeta -1$.
To retrieve the normalization of the spatial correlations from the 2D
function, one needs the information on the radial distribution of the
average galaxy density in the sample.  In the present analysis, the
galaxy concentration varies systematically with the radial distance. We
now derive the formulae relating the  3D CF to the 2D
function and the varying galaxy density.

The 3D CF measures the local average excess of
galaxies relative to the average density of galaxies. Let $\Delta
N_V(r)$ is the average excess of galaxies, i.e. the  number of galaxies
above the average within a volume $\Delta V$ at a distance $r$ from the
randomly chosen galaxy. The CF $\xi(r)$ is
proportional to the number of excess objects:

\begin{equation}
\Delta N_V(r) = \Delta V\, \rho\:\xi(r)\,,
\label{dnv}
\end{equation}

\ni where $\rho$ denotes the average galaxy density. In the present data
the average galaxy density is strongly varying function of the distance
$R$ within each bin, $\rho=\rho(R)$. Using the power-law model for
$\xi$\,:

\begin{equation}
\xi(r) = \left(\frac{r}{r_{\rm o}}\right)^\gamma\,,
\end{equation}

\ni where $r_{\rm o}$ is the correlation length, one can integrate the
galaxy excess equation~(\ref{dnv}) along the line of sight at fixed
transverse separation $p$:

\begin{equation}
\Delta N_A(p) = 
   \Delta A\; G(\gamma)\; r_{\rm o}^{-\gamma}\, \rho(R)\, p{\,^\zeta}\,,
\label{dna}
\end{equation}

\ni where $\Delta A$ is the surface area of volume $\Delta V$ projected
in the sky plane, and $G(\gamma) = \Gamma\left(\frac{1}{2}\right)\,
\Gamma\left(-\frac{\gamma}{2}-\frac{1}{2}\right) \Big/\,
\Gamma\left(-\frac{\gamma}{2}\right)$ \citep{totsuji69}. In the actual
calculations the maximum radial separation of galaxy pairs is limited by
the bin boundaries, what reduces the $G(\gamma)$ factor. The amplitude
of this effect depends on the radial distribution of the galaxy density
(see Sect.~\ref{int_constr}). In the present case, the radial depth of
the distance bin is non-negligible in comparison to the distance $R$
itself.  The counts of galaxy pairs are in fact performed within the
fixed angular separation $\theta$ rather than the $p$. Substituting
$p=\theta\cdot R$ and $\Delta A = \Delta \omega\cdot R^2$ into
equation~(\ref{dna}) we get:

\begin{equation}
\Delta N_\omega(\theta) =
       \Delta \omega\; G(\gamma)\; r_{\rm o}^{-\gamma}\, \rho(R)\;
                \theta^{\,\zeta}\, R^{\,\gamma+3}\,,
\label{dntheta}
\end{equation}

\ni where the subscript $\omega$ indicates that counts are collected
within the solid angle $\Delta \omega$. Notice that here $p$ and
$\theta$ are related to the varying distance $R$ rather than to the bin
centre $R_i$. The excess $\Delta N_\omega(\theta)$ averaged over
all the galaxies found in the selected distance bin is given by:

\begin{equation}
\overline{\Delta N_{\omega\, i}(\theta)} = 
       \Delta \omega\; G(\gamma_i)\; r_{{\rm o}\, i}^{-\gamma_i}\,
                \theta^{\,\zeta_i}\; 
                \frac{\int\, \mathrm d R\, R^{\,\gamma_i+5} \rho_i^2(R)}
                     {\int\, \mathrm d R\, R^2 \rho_i(R)}\,.
\label{dnfinal}
\end{equation}

\ni The index $i$ specifies that only galaxies assigned to the $i$th
distance bin are used to count the galaxy pairs, and $\rho_i(R)$ is the
average density of these galaxies at the distance $R$.  The galaxy
excess at the left-hand side of equation~(\ref{dnfinal}) is equal to that in
equations~(\ref{wI}) and (\ref{power_law}) using the 2D correlations:

\begin{equation}
\overline{\Delta N_{\omega\, i}(\theta)} =
   \Delta \omega\,n_{\rm Io}\,w_i\,\theta^{\,\zeta_i}\,.
\label{2D_3D}
\end{equation}

\ni Here, $n_{\rm Io}$ denotes the observed average surface density of
galaxies of the $i$-th distance bin. The average radial density
distribution, $\rho_i(R)$, in equation~(\ref{dnfinal}) is the actual
distribution that might not be adequately described by the photometric
redshift distribution. The question of statistical reconstruction of the
spectroscopic redshift distribution based on the photometric redshifts
is discussed in the next section.

We use the term `spectroscopic redshift' to denote the perfect distance
measure based on the Hubble expansion. The actual spectroscopic
redshifts are neither sufficiently accurate, nor adequately represent
the Hubble flow. In the present consideration, the statistical
relationships between the redshift distributions is used to assess the
average distribution of galaxy spatial density as a function of
cosmological distance. Thus, small deviations between the `Hubble flow
redshifts' and their spectroscopic counterparts are of no importance.
One should note also that the photometric redshifts are used in the
present calculations exclusively to define distance bins and not to
compute galaxy separations in the radial direction. Thus, the
construction of the 3D CF is not affected by the
`redshift distortions' that modify the separations between objects.

A question of the radial distribution of the average galaxy
density, $\rho(R)$, deserves more comment.  The transverse extension of
the COSMOS field at distances $4000$--$6000$ Mpc amounts to $100$--$150$
Mpc (in comoving units), and even at large distances the survey area is
smaller than typical LSSs. However, both the
transverse survey size as well as the radial extension of the distance
bins are distinctly larger than the galaxy pair separations used in the
CF estimates (nearly all the correlation signal is limited to projected
separations  below $10$\,Mpc). Since the average density $\rho(R)$ and
the CF are determined using the same observational material, the
small-scale fluctuations defined by the CF are measured `on top' of the LSS
potentially present in the data. From that point, it would be desirable to
compare our results with the deep CF assessments in other fields.

\subsection{Modelling the photometric redshifts inaccuracies and failures
\label{zphot}}

\begin{figure}
   \centering
   \includegraphics[width=0.75\linewidth]{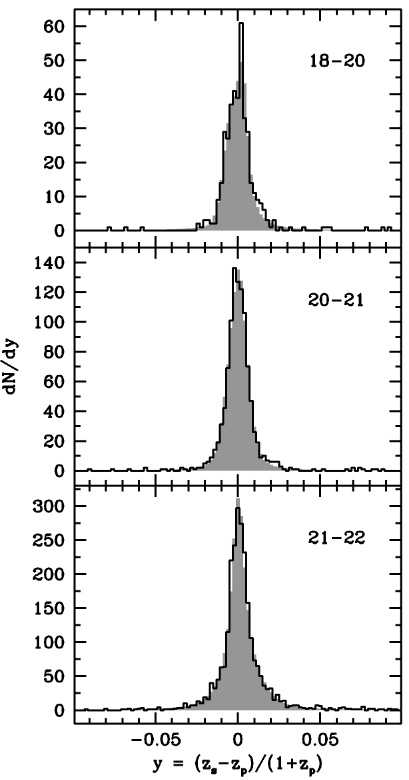}
   \caption{The distribution of $z_{\rm s} - z_{\rm p}$ differences
    in three magnitude bands indicated in the upper right corners.
    Black line histogram -- the observed distribution, the shaded area --
    the ML fit. The ordinate axis shows the number of galaxies with
    known spectroscopic redshifts.}
   \label{zph_fit_1}
\end{figure}

To determine the statistical distribution of spectroscopic redshifts, one
needs to construct the probability distributions of $z_{\rm p}-z_{\rm s}$
differences. Let $p(z_{\rm s}\, |\, z_{\rm p})$ denote the probability
that a galaxy with the assigned photometric redshift $z_{\rm p}$ has
the spectroscopic redshift $z_{\rm s}$. The expected distribution of
$z_{\rm s}$ is then a convolution of the photometric redshifts with the
$p(z_{\rm s}\, |\, z_{\rm p})$ probability:

\begin{equation}
n(z_{\rm s}) =
    \int d z_{\rm p}\; p(z_{\rm s}\, |\, z_{\rm p})\: n(z_{\rm p})\,,
\label{zp_zs}
\end{equation}

\ni where $n(z_{\rm p})$ is the observed distribution of the photometric
redshifts. Since the distribution $p(z_{\rm s}\, |\, z_{\rm p})$ depends
strongly on the galaxy magnitude, it is modelled separately in the
consecutive magnitude bands. A thorough discussions of the $z_{\rm s} -
z_{\rm p}$ deviations (e.g. \citealt{ilbert09,dahlen13}) show that $z_{\rm
p}$ errors are efficiently expressed using the parameter $x = (z_{\rm p} -
z_{\rm s})/(1 + z_{\rm s})$. It was found that a single Gaussian function
adequately reproduces the probability distribution $p(x)$ for the small
absolute values of $x$, i.e. for the `successful' $z_{\rm p}$ estimates.
But the $p(x)$ has broad wings, inconsistent with the narrow central
Gaussian peak, and for still larger $x$ the probability distribution has
small but quasi-constant amplitude.

\begin{figure}
   \centering
   \includegraphics[width=0.75\linewidth]{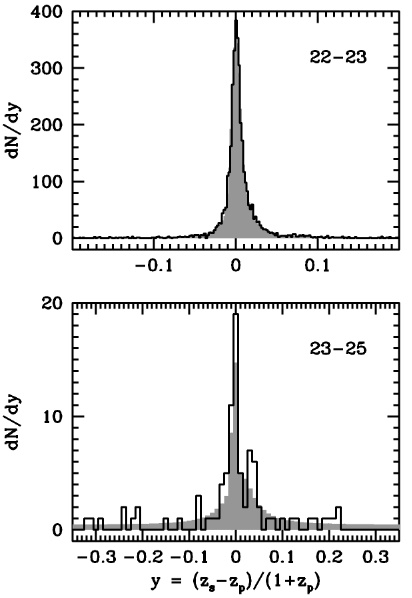}
   \caption{Same as Fig.~\ref{zph_fit_1} for magnitude bands $22-23$ and
    $23-25$.}
   \label{zph_fit_2}
\end{figure}

The objective of our calculations is to assess the $z_{\rm s}$
distribution using the $z_{\rm p}$ data. It is natural to use in
equation~(\ref{zp_zs}) somewhat different parametrization of the $z_{\rm
s}-z_{\rm p}$ differences, viz. $y=(z_{\rm s}-z_{\rm p})/(1 + z_{\rm
p})$. Since even at the fainter magnitudes the photometric redshifts
provide statistically robust estimate of galaxy distances, the
probability distributions $p(y)$ and $p(x)$ have similar construction.
Consequently, function $p(y)$ also exhibits narrow central peak,
relatively wider wings and weak constant signal. To match these
characteristics, we fit the analytic function that is a sum of three
components: (a) the Gaussian -- emulating the narrow peak near $y \approx
0$, (b) the resonance curve -- to reproduce the contribution of larger
$z_{\rm p}$ deviations, and (c) the flat signal -- reproducing the
catastrophic errors. In total, six parameters of the $p(y)$ distribution
were determined. We applied the maximum likelihood (ML) estimation
method. The detailed description of the fitted function
and the whole procedure as well as the numerical results are presented
in Appendix~\ref{app:redshifts}.

The spectroscopic redshifts come mostly from zCOSMOS Data Release
\citep{lilly07}.  We searched also the NED Database, and several
redshifts have been extracted from \citet{onodera12}, \citet{bezanson13} and
\citet{vandesande13}.  The fits are shown in Figs.~\ref{zph_fit_1} and
\ref{zph_fit_2}, where the ordinate axis gives the number of objects
rather than the probability. The galaxy sample was divided in $5$
mag bins.  In agreement with the \citet{ilbert09} discussion, the
fits for galaxies brighter than $23$ mag are strongly peaked at $y = 0$,
while the fainter objects exhibit substantially larger scatter.
Nevertheless, despite a non-negligible number of `catastrophic' $z_{\rm
p} - z_{\rm s}$ discrepancies below $23$\,mag, the central concentration
is still dominant. It allows us to use effectively equation~(\ref{zp_zs}).
However, a relatively small number of the spectroscopic redshifts in the
$23$--$25$ mag bin generates the easily visible noise and broadens the
uncertainty limits of the fitted parameters.  This question is discussed
in detail below.

\subsection{Correlation length \label{corr_length}}

The spatial density of galaxies assigned to the $i$th distance class,
$\rho_i(R)$, in equation~(\ref{dnfinal}) is related to the number of galaxies
$n_i(z_{\rm s})$ in a standard way:

\begin{equation}
\rho_i(R)\; =\; \frac{1}{\Omega\, R^2}\;\frac{{\rm d} n_i(z_{\rm s})}{{\rm d}R}\,,
\label{rho-n}
\end{equation}

\begin{figure}
   \centering
   \includegraphics[width=1.00\linewidth]{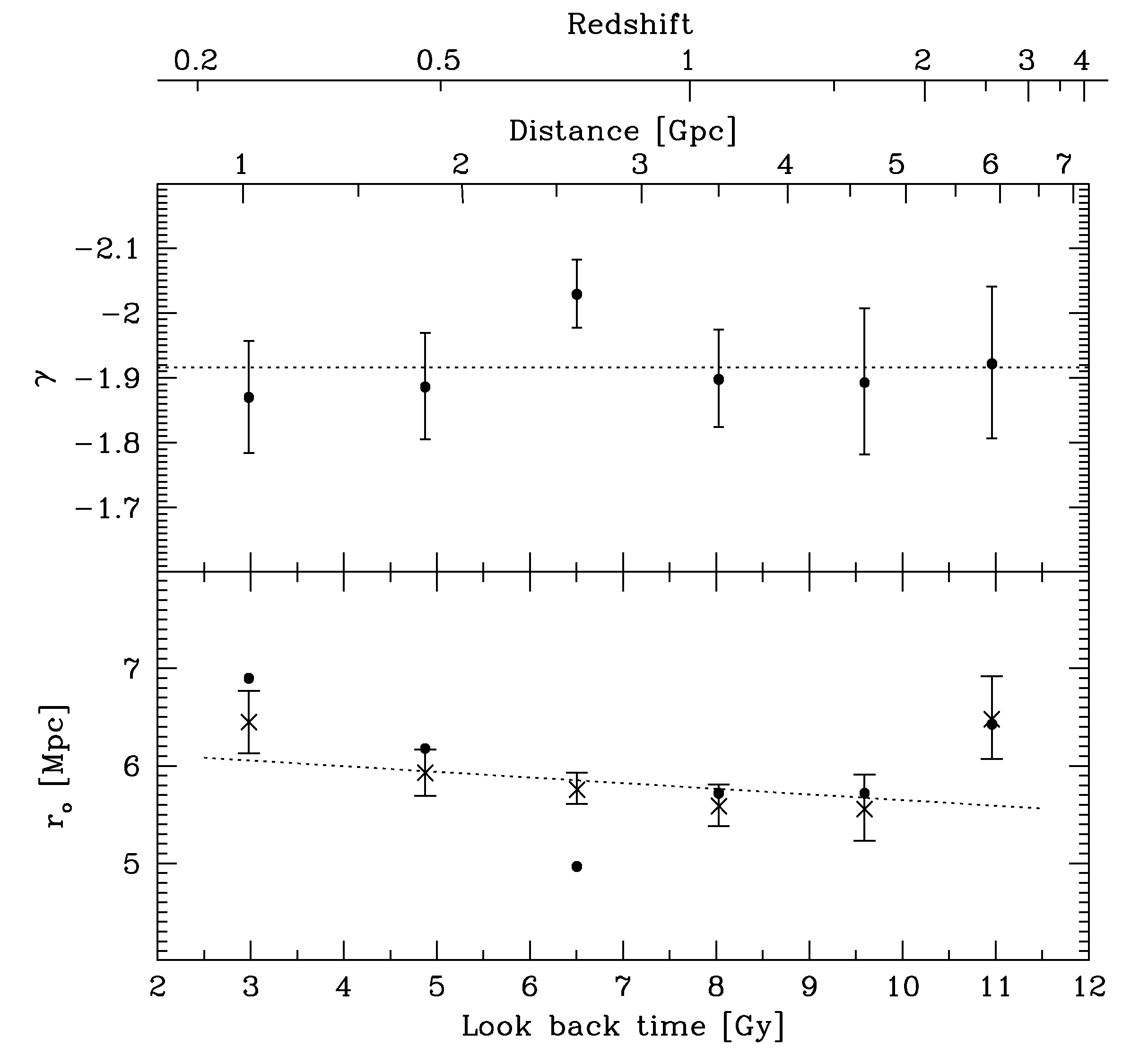}
   \caption{Power-law parameters: $\gamma$ -- slope, and $r_{\rm o}$
    -- normalization, of the spatial correlation function of class A
    galaxies versus look back time. Full dots show simultaneous fittings
    of $\gamma$ and $r_{\rm o}$; crosses in the lower panel are the
    $r_{\rm o}$ best estimates assuming the slope fixed at the average
    value $\gamma = -1.92$.}
   \label{rc_gamma_1}
\end{figure}

\ni where $\Omega$ is the solid angle of the survey. Set of
equations~(\ref{dnfinal})--(\ref{rho-n}) allows us to determine the spatial density
correlation parameter $r_{\rm o}$ for each distance bin. The galaxies in
a given bin are spread over a range of magnitudes. Therefore, the
probability distribution $p(z_{\rm s}\,|\,z_{\rm p})$ in equation~(\ref{zp_zs})
is weighted accordingly to the magnitude distribution of $n_i(z_{\rm p})$
galaxies.  Fig.~\ref{rc_gamma_1} shows our best estimates of the
CF parameters for class A galaxies. The error bars here
are generated solely by uncertainties in the fitting of the power law to
2D CFs (Fig.~\ref{w2_all}). The errors represent
$68$\,\% uncertainties assuming one interesting parameter
(\citealt{avni76}).

A reasoning that photometric redshifts contain sufficient information to
derive the amplitude of the space CF has been  examined with positive
results in the present investigation. Using extensive simulations, we
generated the synthetic galaxy distribution of known space CF and then
effectively determined the CF parameters according to the prescription
presented above. The computational details are described in
Appendix~\ref{app:estimators}.

\begin{figure}
   \includegraphics[width=\linewidth]{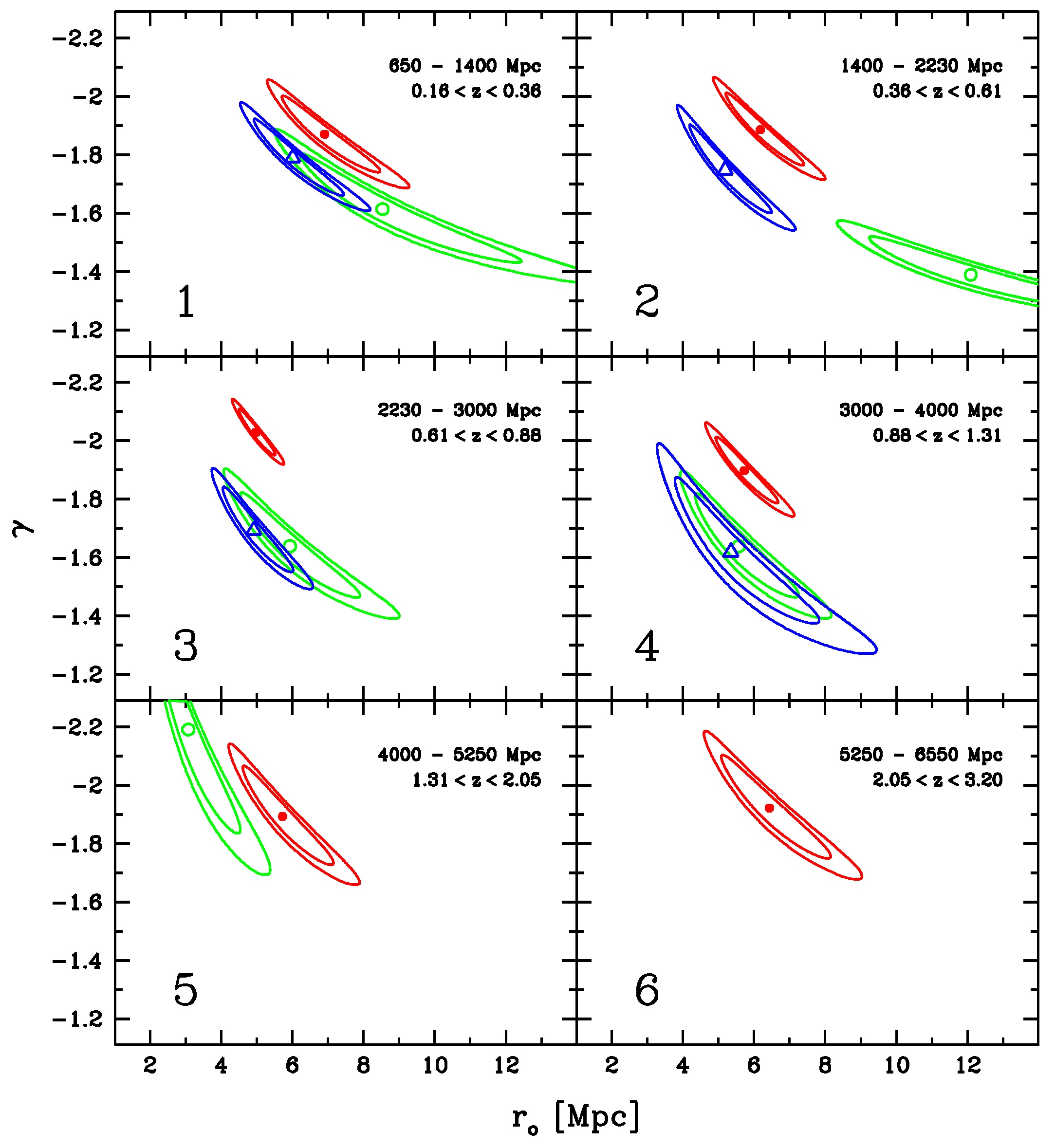}
   \caption{Power-law parameters of the correlation functions in the
   distance bins $1$-$6$ (labelled in the bottom left corner);
   distance/redshift boundaries are indicated in the top right corner.
   Full dots -- class A, open circles -- class B, triangles -- class C.
   Contours show regions of $68$ and $90$ per cent confidence level.}
   \label{rc_gamma_2}
\end{figure}

The simultaneous fitting of $\gamma$ and $r_{\rm o}$ induces strong
correlation of both parameters.\footnote{To be precise, we fit
simultaneously three parameters $\eta_i$, $w_i$, and $\zeta_i$ of
equation~(\ref{basic}). Two of them are `interesting', viz. $w_i$ and $\zeta_i$.
Observational correlation between $w_i$ and $\zeta_i$ is transferred into
correlation between $r_{\rm o}$ and $\gamma$ via equations~(\ref{dnfinal}) and
(\ref{2D_3D}).} In Fig.~\ref{rc_gamma_2}, the $\gamma-r_{\rm o}$ confidence
regions are shown in all the distance bins for galaxies of class A; the
data for the classes B and C are limited to $5$ and $4$ nearest bins,
respectively. The contours do not incorporate uncertainties introduced by
the estimates of the distribution of the $z_{\rm s} - z_{\rm p}$
differences, or the ML fits of $p(y)$. To assess the $p(y)$ uncertainties,
the mock $p(y)$ distributions were created using the bootstrap method. The
details of the whole procedure are described in Appendix~\ref{app:redshifts}.

We limited our analysis to the faintest galaxies because at the brighter
magnitudes the $p(y)$ distributions are based on an extensive data and
the uncertainty of the $p(y)$ fits is negligible. Small number of
spectroscopic redshifts at faint magnitudes results in the poor quality
of the $p(y)$ fit in the $23 < i^+ < 25$ band. The galaxies in the
$23$--$25$ mag dominate in the distance bins $4 - 6$. They constitute
$71$, $97$ and $99.5$\,\% of the class A galaxies in the distance bins
$4$--$6$, respectively.

The calculations of $r_{\rm o}$ for the simulated $p(y)$ probability
functions were repeated as for the actual data. We find that the rms
scatter of $r_{\rm o}$ produced by the uncertainties of the ML fitting
amounts to $0.16$, $0.26$ and $0.29$\,Mpc for bins $4$--$6$. Thus, the
errors related to the ML fits are several times smaller than the
uncertainties defined by the confidence level contours. Assuming that
both errors add in squares, the uncertainties involved in the ML fitting
do not contribute significantly to the total uncertainties.

\subsection{The radial extension of distance bins \label{int_constr}}

\begin{figure}
   \includegraphics[width=1.00\linewidth]{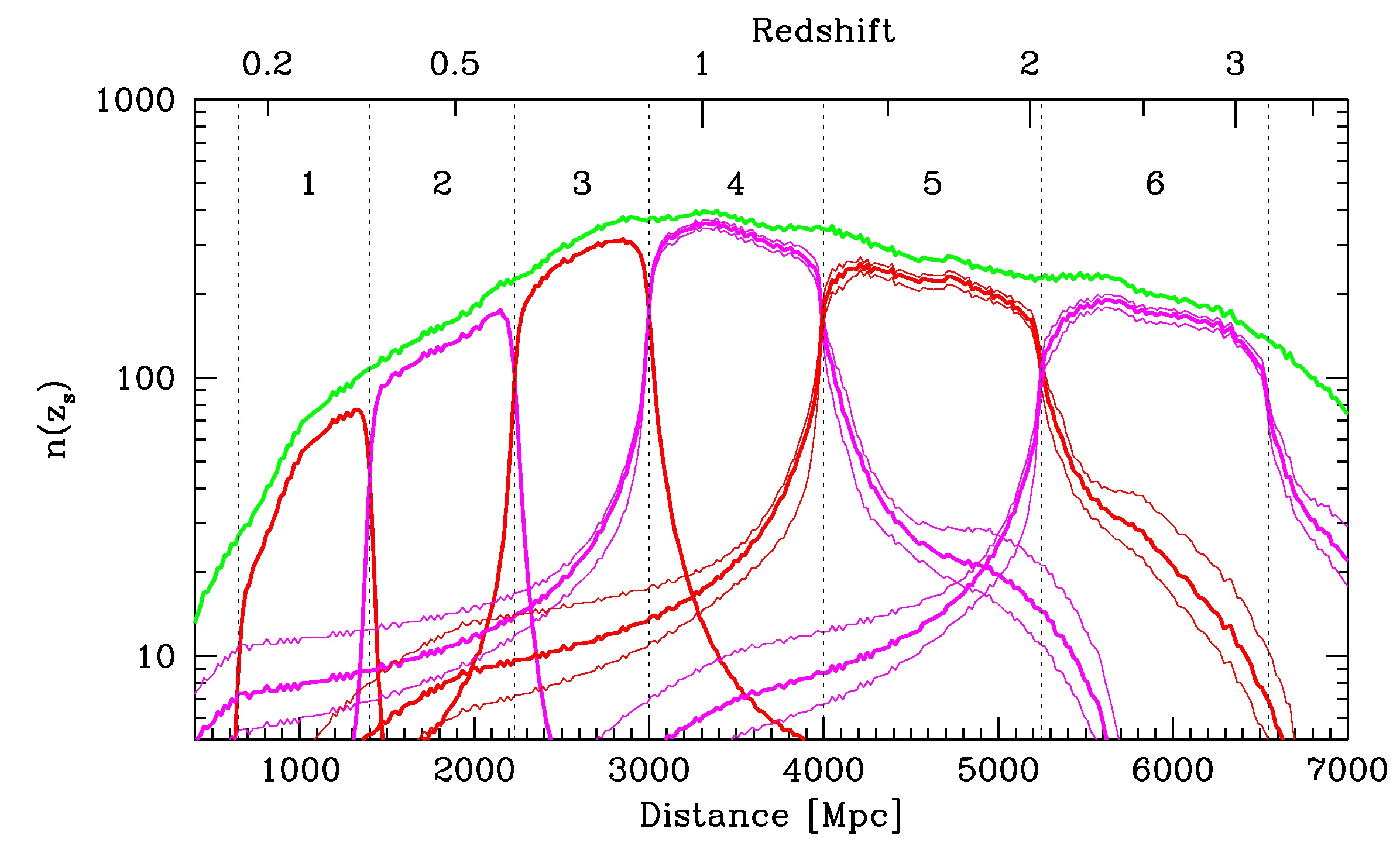}
   \caption{Thick curves -- numbers of class A galaxies in six
   distance bins (indicated by labels $1$--$6$) corrected for the
   $z_{\rm p}$ errors using equation~(\ref{zp_zs}). Thin curves -- $1\sigma$
   uncertainty regions resulting from uncertainties of the 
   $z_{\rm s} - z_{\rm s}$ fits (see Appendix A for details.); shown
   only for three most distance bins where the errors are more
   pronounced. The upper envelope shows the summed galaxy numbers. The
   distance bins partially overlap because galaxies are assigned to
   their bins based on the photometric redshift.}
   \label{bin_dens}
\end{figure}

In Fig.~\ref{bin_dens}, we plot the radial distribution of the space
density of galaxies luminosity class A calculated using equations (\ref{zp_zs})
and (\ref{rho-n}). Although the effects of the photometric redshift errors
are easily visible, the bins are still well defined in the real space.
Long tails of the three most distant bins stretching towards the
lower distances result from the catastrophic $z_{\rm p}$ errors.

It was shown in the previous section that the uncertainties of the
$p(y)$ fits contribute marginally to the total errors of our $r_{\rm o}$
estimates. This is because the $p(y)$ scatter only weakly affects the
$n(z_{\rm p}) \rightarrow n(z_{\rm s})$ transformation. The thin curves
for three most distant bins in Fig.~\ref{bin_dens} show the rms
uncertainty range of our $n(z_{\rm s})$ reconstruction produced by the
stochastic character of the $p(y)$ estimates. It is visible that
statistical uncertainties strongly affect only the low-amplitude tails
of the individual $n(z_{\rm s})$ distributions.
 
The composite shape of the $\rho_i(R)$ distribution is accounted for in
the calculations of the correlation length $r_{\rm o}$. One should notice
that the contribution to the $r_{\rm o}$ amplitude of the low-density
extensions (that suffer from large uncertainties) is small as compared to
the `central' section of the the bin. This is because the integral at
the right-hand side of equation~(\ref{dnfinal}) contains the density square. 

Relationships between the measured 2D CF and the 3D CF derived using
equations~(\ref{dnfinal}) and (\ref{2D_3D}) should be corrected for the finite
radial extensions of bins. However, because the bin depths are
substantially larger than the expected maximum separations at which the
CF significantly deviates from zero, the effect is small as compared to
statistical uncertainties. It was assessed as follows.

Counting galaxy pairs in the chosen distance bin corresponds to the
double integration along the line of sight of the 3D CF weighted by the
actual radial galaxy space density distribution. The difference between
the results based on the equation~(\ref{dna}) and the pair number counts
depends on the a priori unknown parameters of $\xi(r)$.  Since this
difference is expected to be small in the present calculation, we
expressed the amplitude of the finite bin depth effect a posteriori,
taking the best-fitting $\xi(r)$ parameters as determined in the previous
sub-section.  In all the bins, the relative difference between the
analytic value used in equation~(\ref{dna}) end the `effective' $G$-factor is
negligible at the small separations, rises to $\sim 1$ per cent at the
transverse separations of $1$--$3$\,Mpc, and reaches $5$--$8$ percent
(depending on the bin number) at $\sim 20$ Mpc. The expected impact of
this `integral constraint' on the CF slope $\gamma$ would be of the
order of $\sim 0.02$ with the correspondingly small effect for the
correlation length.


\section{Discussion and conclusions}
\label{sec:conclusions}

The mean absolute magnitude of galaxies in the sample A coincides roughly
with the characteristic magnitude $M^\star$ in the Schechter luminosity
function.  Because of the flux selection, the mean magnitude difference
between the samples A--C varies with the distance. In the first three
distance bins, galaxies in the sample B are fainter than class A galaxies by
$\sim 1.4$\,mag, while in the distance bins $4$ and $5$ this difference
drops to $1.3$ and $1.2$, respectively. The magnitude separation between
sample A and C is more strongly affected by the selection. It amounts to
$2.9$\, mag in the first two distance bins, and is reduced to $2.7$ and
$2.3$ in the next two bins, respectively.

A comparison of our CF slope estimates with a number of
fragmentary measurements present in the literature leads to somewhat
ambiguous conclusions.  \citet{marulli13} compile a number of recent
results on the CF parameters derived for a wide range of
galaxy luminosities and redshifts. Our results fit well to the general
distribution of measurements in their fig.~5. In particular, the slope
flattening in the sample B and C as compared to the sample A seems to be
present also in the published results (e.g.  \citealt{pollo06,coil06}).
One should note, however, that individual measurements refer frequently to
different redshifts and are subject to large uncertainties. Therefore,
questions on the specific relationships between CF
parameters still remain open. The VIPERS data (\citealt{marulli13}) that
suffer from the smallest uncertainties, indicate a weak but systematic
flattening of the slope with redshift, while this effect is not indicated
by our investigation. Furthermore, these data show only a marginal
CF slope -- absolute magnitude dependence.
One should note, however, that the VIPERS galaxies in the \citet{marulli13}
data span a relatively narrow magnitude range of $\Delta M \approx 1.5$,
and are confined to redshifts between $0.5$ and $1.1$.

Despite considerable size of the banana shape confidence regions in
Fig.~\ref{rc_gamma_2}, our estimates of $\gamma$
and $r_{\rm o}$ provide constraining information on the cosmic evolution
of the galaxy CF.  Although, the relative positions of
the best-fitting parameters in the $r_{\rm o}$--$\gamma$ plane of class
A--C galaxies vary from one distance bin to the another, the shape and
orientation of the confidence regions demonstrate some permanent
characteristics of the CF over a wide redshift range.
It appears that the correlation length, $r_{\rm o}$, of class A--C
galaxies is quite similar in most of the distance bins, although not
necessarily identical. Larger discrepancies between class B and A in the
distance bin $2$ and $5$ are of the opposite sign, and seem to be
generated by statistical fluctuation in the data rather than the cosmic
signal. The conclusion in the previous section that the CF
slope is steeper for the most luminous galaxies (class A) is
strengthen by the overall layout of the confidence regions.

\citet{marulli13} reach the opposite conclusion. They find `a monotonic
increase in the clustering length $r_{\rm o}$' as a function of magnitude
`in all three redshift ranges considered', i.e. $0.5$--$0.7$, $0.7$--$0.9$
and $0.9$--$1,1$. Taking into account elongated shapes of the confidence
regions, such interpretation of Fig.~\ref{rc_gamma_2} is not completely
ruled out; however, it is not favoured by the present results.

It is instructive to compare our results with the \citet{arnalte-mur14}
investigation, as it also uses photometric redshifts, and covers a
relatively wide redshift range, $0.35 < z < 1.25$. Their fig.~7
apparently indicates a rise of the correlation length with galaxy
luminosities, but no obvious dependence of the correlation slope on the
luminosities. Despite these differences, both investigations support the
conclusion that the clustering amplitude increases with the galaxy
luminosities. Our Fig.~\ref{rc_gamma_2} shows that the present
calculations are not immune to statistical fluctuations.  The question
of systematic errors is more ambiguous. It is assumed in the present
method that all the known as well as unrecognized selection effects that
alter the galaxy distribution are accounted for using the information
incorporated in the catalogue itself. In the alternative methods, the
mock catalogues are used as a reference `random' distribution. Such
catalogues are designed to model all the instrumental/observational
constraints that modify the genuine galaxy clustering. A question which
of these two methods is more suitable for dealing with various
observational biases depends on the particular properties of the
observational data. The overall characteristics of the COSMOS/Subaru
catalogue, namely very wide redshift coverage of relatively small
surface area with numerous gaps, favour the method developed in this
paper.

Because of the apparently constant slope of the CF for
the class A galaxies, it is reasonable to fix the slope for all the bins
at the average value of $\overline{\gamma} = -1.92$.  The best estimates
of $r_{\rm o}$ in this case are shown with crosses in the lower panel of
Fig.~\ref{rc_gamma_1}. There is a weak indication that $r_{\rm o}$
slowly increases with the cosmic time.  The dotted line in the lower
panel shows the least-squares linear fit to the $r_{\rm o}$ versus look back
time. The slope of this line differs from $0$ at $1.14\,\sigma$.

The issue whether the present results are representative for the
general galaxy clustering properties or refer just to the COSMOS field
cannot be answered using this field alone. The question is legitimate
because of some peculiarities in the field were reported in the
literature (e.g. \citealt{meneux09}). We would postpone the answering to
this until the similar analysis is performed on other deep fields. 

\subsection{The bias}

The homogeneous computation scheme applied to the galaxies spread over
a huge redshift range allowed us for a uniform assessment of the
clustering evolution as well as the galaxy bias, $b(z)$:

\begin{equation}
b(z) = \sigma_{\rm g}(z) / \sigma_{\rm m}(z)\,,
\end{equation}

\ni where $\sigma_{\rm g}(z)$ denotes the galaxy number rms fluctuations
and $\sigma_{\rm m}(z)$ the mass rms fluctuations within the same volume.
The CF is rigorously connected to the galaxy rms
fluctuations. In particular, the rms of the galaxy number in a sphere
of radius $r$ is related to the power-law CF (\citealt{peebles76}):

\begin{equation}
\sigma_{\rm g}(r|z) = [ C_\gamma\; \xi(r|z) ]^{1/2}\,,
\end{equation}

\ni where $C_\gamma = 72\cdot 2^\gamma
                    /[(3+\gamma)(4+\gamma)(6+\gamma)]$, and
the volume size as well as
the redshift dependence of $\sigma$ and $\xi$ are explicitly indicated.
The growth of the matter fluctuations, $\sigma_{\rm m}(r|z)$,
in the linear regime is completely defined by the cosmological model.
In the following, we use the set of equations listed by \citet{meneux08}:
$\sigma_{\rm m}(r|z) = \sigma_{\rm m}(r|0) / D(z)\,$,
\ni where $D(z) = g(z)/[g(0)\,(1+z)]$ and
$g(z) = \frac{5}{2}\,\Omega_{\rm m}\, \Big/ \left[ \Omega_{\rm m}^{4/7} -
        \Omega_\Lambda + (1+ \Omega_{\rm m}/2)\,(1 +
        \Omega_\Lambda/70)\right]$
\ni is the normalized growth factor \citep{carroll92}. The evolving
density parameters $\Omega_{\rm m}$ and $\Omega_\Lambda$ are related to theirs
present epoch values:

\begin{equation}
\Omega_{\rm m} \equiv \Omega_{\rm m}(z) =
    \frac{\Omega_{\rm m,o}\,(1+z)^3}{E^2(z)}\,, \hspace{2mm}
\Omega_\Lambda \equiv \Omega_\Lambda(z) =
    \frac{\Omega_{\Lambda,{\rm o}}}{E^2(z)}\,,
\end{equation}

\begin{figure}
   \centering
   \includegraphics[width=1.00\linewidth]{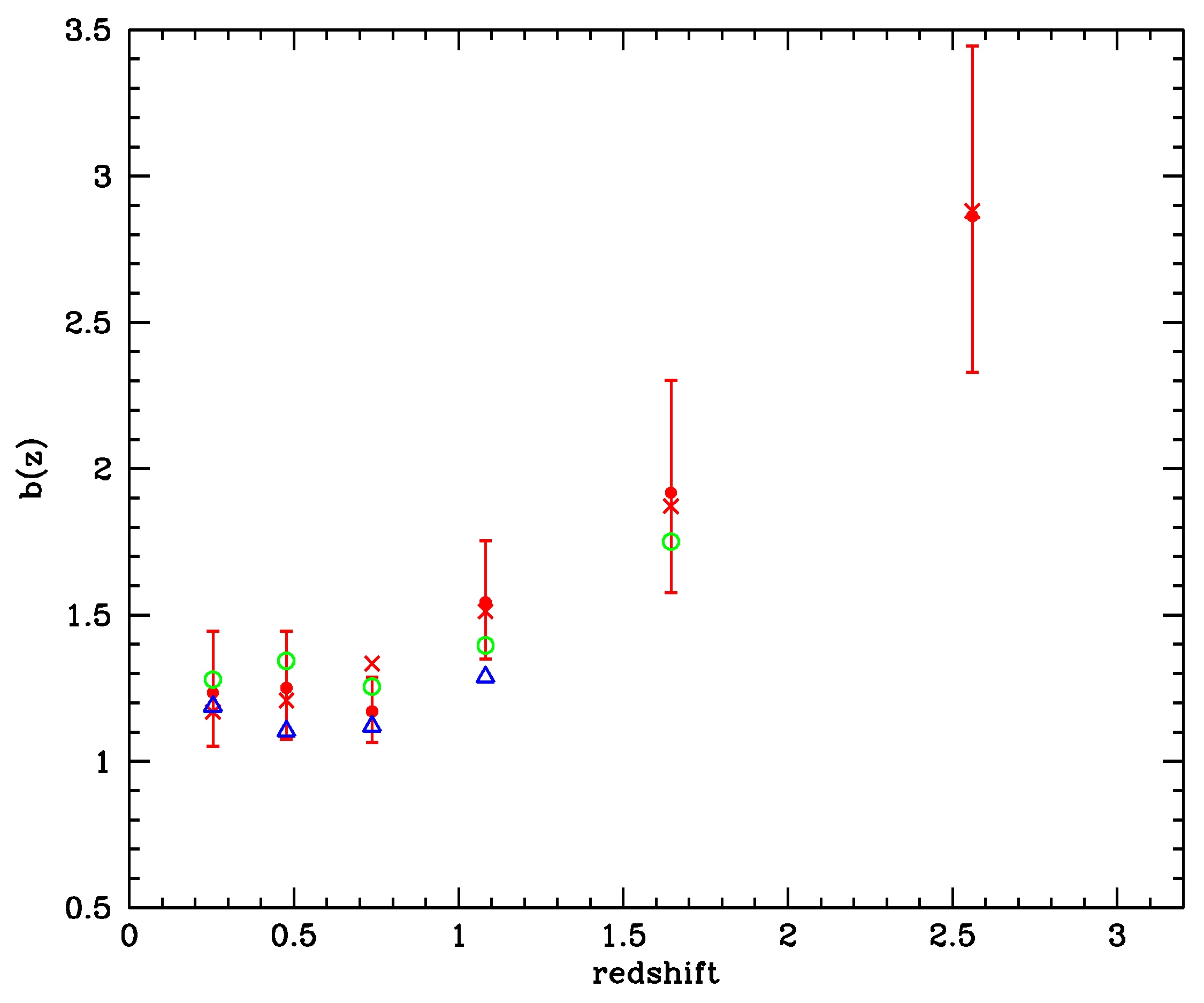}
   \caption{The redshift evolution of the linear bias factor.
   Dots denote the bias factor calculated for the $r_{\rm o}$ and
   $\gamma$ fits of class A galaxies indicated with dots in
   Fig.~\ref{rc_gamma_1}. The error bars correspond to $68$\,\%
   uncertainties drawn in Fig.~\ref{rc_gamma_2}.
   Crosses show the bias of the class A galaxies assuming the slope
   of the correlation function fixed at $\gamma = 1.92$ (crosses in
   Fig.~\ref{rc_gamma_1}). The bias factor for B and C classes are
   shown with open circles and triangles, respectively.}
   \label{bias}
\end{figure}

\ni where $E^2(z) = \Omega_{\Lambda,{\rm o}} + 
                    \Omega_{\rm m,o}\,(1+z)^3$\, is the expansion factor
for the flat cosmological model, for which
$\Omega_\Lambda + \Omega_{\rm m} = 1$.
The present epoch mass fluctuations amplitude, $\sigma_{\rm m}(0)$, is
commonly normalized to the mass rms within a sphere of radius
$r=8\, h^{-1}$\,Mpc, where $h=H_{\rm o}/100$\,km\,s$^{-1}$Mpc$^{-1}$.
We use the \citet{komatsu11} figure $\sigma_{\rm m}(0) = 0.82$ based
on $7$ yr {\it WMAP} observations. The bias--redshift relations
deduced from the CF measurements are shown in
Fig.~\ref{bias}. Dots and crosses indicate the bias variations of
the class A galaxies, circles -- class B, and triangles -- class C.
Growth of $b(z)$ for the most luminous (class A) galaxies reported
in the literature for low and moderate redshifts (e.g.
\citealt{marinoni05,papageorgiou12}) continues to high redshifts.
The data for galaxies class B and C are limited to low redshifts and
apparently the bias amplitude does not depend on the galaxy luminosities
in that area. It is difficult to assess whether the bias coincidence
at low redshift of the class A--C galaxies contradicts the well documented
relationship between the bias and galaxy luminosities. This is because
a sharp rise of the bias amplitude with galaxy luminosities
(e.g. \citealt{zehavi11}) is mostly constrained to galaxies brighter
than  $M^\star$, while our class A corresponds roughly to $M^\star$,
and classes B and C are fainter.

In low and moderate redshifts the present results are largely consistent
with the measurements based on the galaxy samples with spectroscopic
redshifts. The photometric redshifts provide only the statistical
constraints on the galaxy distribution, and cannot be used to localize
precisely individual objects. This inevitably broadens the uncertainties
of our estimates.  However, the photometric redshifts can be determined
in massive scale over much larger volume of the universe than it is now
feasible for the spectroscopic surveys. The present calculations are
based purely on the observational data, i.e. in the calculations we do
not include mock galaxy catalogues conceived from the cosmological
simulations. This additionally increases our final uncertainties.
However, our method is more adequate as long as effects of the cosmic
variance on the galaxy clustering are not fully accounted for.

Our high-redshift measurements of the galaxy linear bias fit well in the
low-redshift limit to the previous studies. To assess more precisely
the evolution trends of the galaxy CF at high
redshifts, we plan to perform analogous investigation on other deep
galaxy photometric redshift surveys.

\section*{acknowledgements}

We thank the anonymous reviewer for her recommendations
that greatly helped us to improve the presentation of our analysis.
Based on zCOSMOS observations carried out using the Very Large Telescope
at the ESO Paranal Observatory under Programme ID: LP175.A-0839.
This work has been partially supported by the Polish
NCN grant 2011/01/B/ST9/06023.

\bibliography{soltan_sh}

\appendix

\section[]{Analytic fits of the \lowercase{\boldmath $z_{p} - z_{s}$}
deviations}
\label{app:redshifts}

\begin{table*}
 \centering
 \begin{minipage}{150mm}
  \caption{Parameters for the $p(y)$ distributions in 5 mag bins.}
  \begin{tabular}{lrrrrrrrr}
  \hline
  \multicolumn{1}{c}{Mag bin}&$N_{\rm s}$\footnote
   {Number of galaxies with measured spectroscopic redshifts used in the fittings.}
                      &$\alpha_{\rm o}$&
                              $\alpha_{\rm g}$&
                                          $\mu_{\rm g}$
                                                 &$s_{\rm g}$ &
                                                         $\alpha_{\rm r}$
                                                                    &$\mu_{\rm r}$ 
                                                                             &$s_{\rm r}$ \\
  \hline
$18$--$20$&   407    & 0.0028 & 0.3368 & -0.0045 & 0.0042 & 0.6604 & 0.0028 & 0.0039 \\
$20$--$21$&  1050    & 0.0085 & 0.5423 & -0.0006 & 0.0059 & 0.4492 & 0.0009 & 0.0049 \\
$21$--$22$&  2521    & 0.0120 & 0.1942 &  0.0037 & 0.0110 & 0.7938 & 0.0004 & 0.0045 \\
$22$--$23$&  3775    & 0.0239 & 0.3300 & -0.0008 & 0.0044 & 0.6461 & 0.0039 & 0.0094 \\
$23$--$25$&    89    & 0.1773 & 0.1886 & -0.0034 & 0.0026 & 0.6341 & 0.0042 & 0.0298 \\
  \hline
  \multicolumn{9}{c}{Simulations} \\
$23$--$25$&  Average & 0.2090 & 0.2482 & -0.0037 & 0.0036 & 0.5428 & 0.0096 & 0.0256 \\
           &  rms     & 0.0968 & 0.1110 &  0.0015 & 0.0033 & 0.1710 & 0.0132 & 0.0138 \\
  \hline

  \end{tabular}
 \end{minipage}
\end{table*}

Distribution of  $z_{\rm p} - z_{\rm s}$ differences  is represented using the
probability distribution $p(y)$, where $y=(z_{\rm s} - z_{\rm p})/(1+z_{\rm p})$.

Complex nature of the $z_{\rm p} - z_{\rm s}$ deviations requires adequate
functional form that accounts for a high fraction of almost perfect
$z_{\rm p}$ measurements as well as casual errors and sporadic failures.
Accordingly, the $p(y)$ distribution is defined as the normalized sum of
three components: the Gaussian function $g(y)$, the resonance function
$r(y)$ and a low amplitude uniform probability $a_{\rm o}$:

\begin{equation}
p(y) = \alpha_{\rm g}\cdot g(y) + \alpha_{\rm r}\cdot r(y) + \alpha_{\rm o}\,,
\end{equation}

\ni where
\begin{equation}
\begin{array}{lcl}
g(y) & \mbox{=} & \frac{1}{\sqrt{2\upi}\, s_{\rm g}}\;\exp[-(y-\mu_{\rm g})^2/2s_{\rm g}^2] \\
r(y) & \mbox{=} & \frac{1}{\upi}\, \frac{s_{\rm r}}{(y-\mu_{\rm r})^2 + s_{\rm r}^2}\,.
\end{array}
\end{equation}

\ni Parameters $\mu_{\rm g}$, $s_{\rm g}$, $\mu_{\rm r}$ and $s_{\rm r}$
plus relative contributions of the components, $\alpha_{\rm g}$,
$\alpha_{\rm r}$, and $\alpha_{\rm o}$
are fitted to the observed distributions of the $y$ parameter. The
ML estimation was applied. Since the quality of the
photometric redshifts deteriorates with increasing magnitude, the data
were divided into $5$ mag bands. A small number of measured
spectroscopic redshifts below $m = 23$ and a high fraction of $z_{\rm
p}$ catastrophic errors potentially could introduce relatively large
uncertainties to the present analysis. The question to what extent
uncertainties related to the fitting procedure affect our estimates of
the space correlation amplitude is addressed as follows.

We generated $200$ quasi-random samples of the $(z_{\rm s}, z_{\rm p})$
pairs using the bootstrap method. The mock samples were drawn
from the real data. The average amplitudes of parameters fitted to the
simulations and their rms scatter have been calculated. The results of
the entire procedure are listed in Table A1. The first five rows give
the best-fiting parameters to the real data while the bottom two rays show
the results for the simulated $23$--$25$ mag bin.

Because of the distinctive form of the fitted function the parameters
are strongly correlated. Therefore the resultant $p(y)$ distributions
exhibit a moderate variations, despite the high scatter of individual
parameters. Stable character of $p(y)$ distribution generates via
equation~(\ref{zp_zs}) well-defined spectroscopic redshift distributions,
$n(z_{\rm s})$.  We assessed the uncertainties of $n(z_{\rm s})$ from a
spread of the mock distributions. A set of $200$ simulated spectroscopic
redshift data, $n(z_{\rm s})$, was generated using the simulated $p(y)$
probability functions. The rms scatter of $n(z_{\rm s})$ is indicated in
Fig.~\ref{bin_dens} with thin curves for three most distant bins.

\section[]{Testing the estimators of CF parameters}
\label{app:estimators}

The full procedure of measuring the 3D CF parameters using the catalogue
of photometric redshifts has been tested on the simulated data. First,
we generate a 3D point distribution according to a priori defined
statistical characteristics. Then, the data are processed in the
same way as the COSMOS observations to retrieve the CF parameters.
Finally, the results are confronted with the original values.

\subsection{Modelling the space distribution with definite CF}

The objective was to construct the model distribution of points
characterized by the power-law CF with the slope and normalization close
to those determined for the COSMOS data. This synthetic material
was not devised to imitate other statistical properties of the
real galaxy population. From among a diversity of space
distributions satisfying the selected CF, we took a computationally
manageable position-dependent function. The particular point
distribution was realized by the MC method. The points were drawn
according to the properly defined probability distribution. The
algorithm that provided an acceptable approximation for the power-law CF
was constructed as follows.

Two populations of points are distributed within a specified volume: (a)
points concentrated in `clusters',\footnote{Here, `clusters' represent
clamps of points  that generate the required CF, with no relation
to the real galaxy clusters.} and (b) `field' points distributed
randomly, but outside clusters. The cluster centres are distributed
fully randomly. Points within each cluster are distributed according to
the broken power law:

\begin{equation}
\label{cl_profile}
\rho(r) = \left\{ \begin{array}{lcl}
          \rho_{\rm b}\, \left( \frac{r_1}{r_2} \right)^{\alpha_2}
                           \left( \frac{r}{r_1} \right)^{\alpha_1} &
                                                        \mbox{for} &
                                                 r < r_1 \medskip \\
            \rho_{\rm b}\, \left( \frac{r}{r_2} \right)^{\alpha_2} &
                                                        \mbox{for} &
                                         r_1 \le r < r_2 \medskip \\
                         \rho_{\rm b} & \mbox{for} & r \ge r_2\,, \\
                   \end{array} \right.
\end{equation}

\ni where $\rho_{\rm b}$ is the space density of points outside
clusters, $r_1$ and $r_2$ are characteristic cluster radii that
delineate two zones of distinct power indices $\alpha_1$ and $\alpha_2$.
The amplitude of the CF depends also on the number of clusters or
equivalently -- on the fraction of volume occupied by clusters. In the
subsequent calculations, we adopted the following parameter values: radii
$r_1$ and $r_2$ of $6$ and $16$\,Mpc, respectively, and the power-law
indices $\alpha_1 = -2.44$ and $\alpha_2= -1.25$ for the central and
outer cluster zones. The volume occupied by clusters was selected at
$\varkappa \approx 0.40$ of the total survey volume. This set of
parameters determines the relative contribution of field and clustered
points to the total number at $0.33$ and $0.67$, respectively.

\begin{figure}
   \centering
   \includegraphics[width=1.00\linewidth]{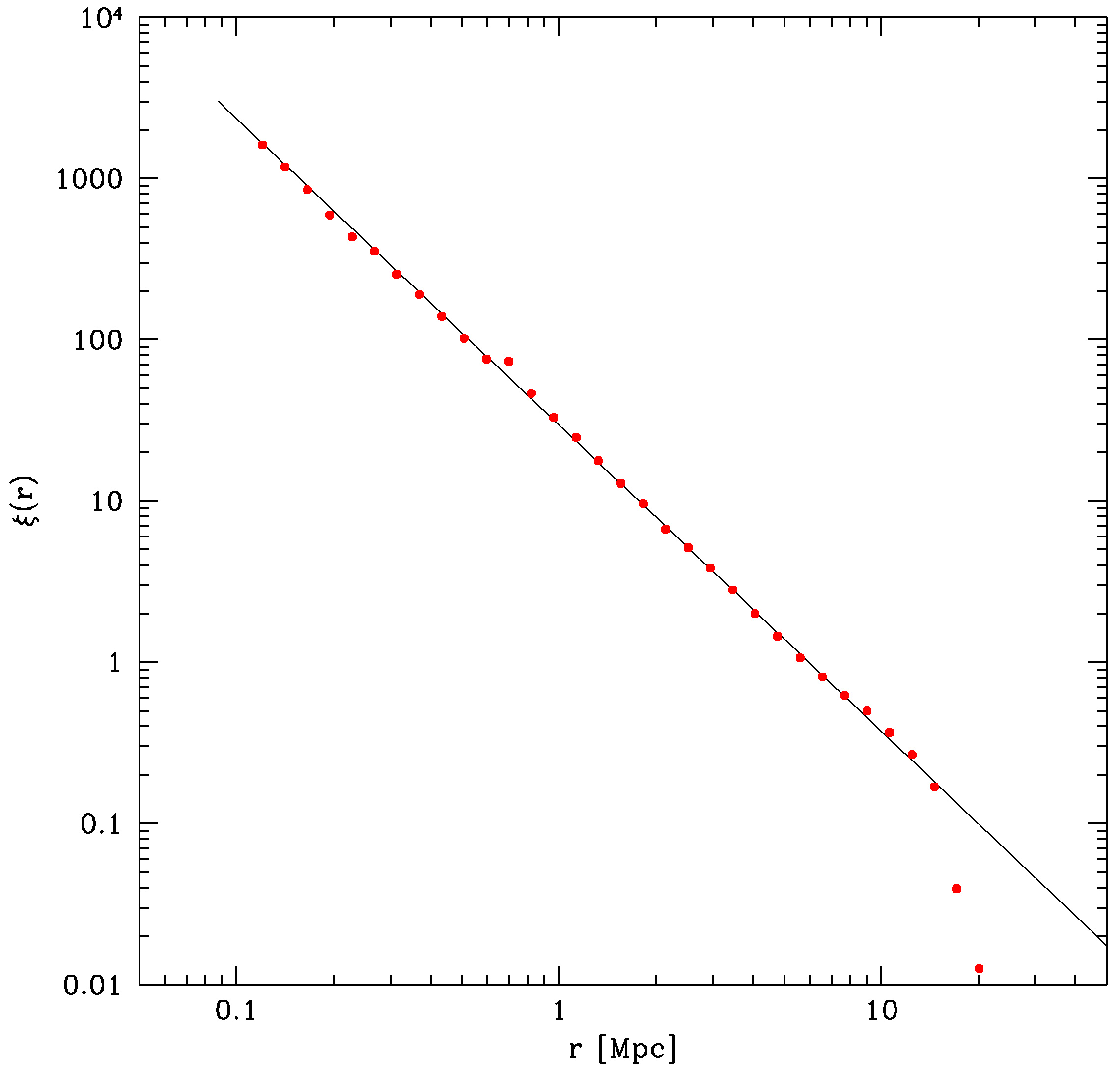}
   \caption{Dots -- the correlation function of $62158$ points
   distributed according to the model described in the text, line --
   the power-law fit $\xi(r)=(r/r_{\rm m})^{\gamma_{\rm m}}$ with
   $r_{\rm m}=5.89$\,Mpc and $\gamma_{\rm m} = -1.89$.}
   \label{app:model_acf}
\end{figure}

The present prescription of the space point distribution yields the CF
emulating the power law over a wide range of separations extending up to
$\sim 15$\,Mpc. This is illustrated in Fig.~\ref{app:model_acf}, where $\sim
62000$ points were distributed in a volume of $1.73\times 10^7$\,Mpc$^3$
according to equation~(\ref{cl_profile}) with $\rho_{\rm b} =
2.25\times\nolinebreak 10^{-3}$\,Mpc$^{-3}$. Fitting the power law to the
CF generated by this set of parameters gives the slope $\gamma_{\rm m} =
-1.9$ and the correlation length $r_{\rm m} = 5.9$\,Mpc.

It is important to emphasize that this model distribution generates the
definite CF but has no resemblance to the actual galaxy space
distribution. The objective of the procedure is just to test
the efficiency of the method, i.e. whether the algorithm presented in
Sections~\ref{sec:2Dacf} and \ref{sec:3Dacf} is able to extract the
parameters of the CF from the surface data. 
One can expect that individual realizations of the synthetic model
limited to the volume of the COSMOS field would provide the CF 
parameters scattered around the original values assumed in the
simulations. The size of this spread is characteristic to the model
and does not represent uncertainties of the CF parameters derived
from the real data.

\subsection{Extracting the PL parameters from the model data distributed
in the COSMOS field}

We now adjust our model to the global constraints that shape the COSMOS
data. The objective is to reproduce some basic statistical
characteristics (such as the total number of objects and a form of the
radial distribution) of the most luminous galaxies, denoted as class A.

First, the points are distributed in a pyramid volume of the solid
angle of $1.42\!\times\! 1.38$\,deg$^2$, what corresponds to the size
of the COSMOS/Subaru field. To emulate a decreasing space density of
galaxies at large distances, the probability of qualifying a  point for
further processing was adequately reduced. In effect, the final set of
`objects' contained $\sim 76000$ points.  We declare that the present
method is essentially insensitive to the numerous empty regions
distributed over the field as well as to the smooth, large angular
scale inhomogeneities of the survey. Although variations of the
limiting magnitudes have been investigated (\citealt{taniguchi07}), and
the final catalogue is free from substantial inhomogeneities, one
should allow for some residual instrumental fluctuations that simulate
the clustering signal. To check that, we ran the tests with a
superimposed mask that imitates holes in the actual catalogue due to the
bright stars. The `Swiss cheese' mask contains above $125$ circles
representing the most prominent blank fields. Additionally, we test the
effects of potential angular inhomegeneities of the survey. This is
achieved by applying the `fluctuation filter', F$_a$. We superimpose
a filter that introduces 
fluctuations of the point surface density with a characteristic
wavelength $\lambda = 40$\,arcmin in both directions (RA and Dec.):

\begin{equation}
n(x,y) = \overline{n}\, [1 + a\, \sin(kx)\, \sin(ky)]\,
\end{equation}

\ni where $n(x,y)$ and $\overline {n}$ are the local and the mean point densities,
respectively, $a$ is the fluctuation amplitude, and $k= 2\upi /40\,$arcmin.
Three amplitudes  $a$ were applied: $0.025$, $0.05$ and $0.10$ 

`Spectroscopic redshifts' were assigned to all the objects assuming the
cosmological model adopted in the paper. To attach a `photometric
redshift' to each object, we apply a method identical to that described
in Appendix~\ref{app:redshifts}. Using the ML method, we find the
best-fitting parameters of the $p(x)$ probability distribution, where
$x=(z_{\rm p} - z_{\rm s})/(1+z_{\rm s})$. Since the  overall shape of
$p(x)$ and $p(y)$ is similar, the functional form of both distribution
is the same. The COSMOS $p(x)$ distributions depend on the apparent
magnitude of objects. To mimic this effect, each object in our catalogue
was labelled with the `apparent magnitude' drawn from the magnitude
distribution of the real data. 

The final simulated catalogue contained the list of points identified just by
their angular coordinates and photometric redshifts. Such `observational
material' was analysed using the method and formulae in
Sections~\ref{sec:2Dacf} and \ref{sec:3Dacf}. Calculations were performed for
$25$ sets of simulated catalogues, using various combinations of the bright
star mask (BS) and fluctuation filters (F$_{0.025}$, F$_{0.05}$ and
F$_{0.10}$).  The quantitative comparison of the CF parameter estimates,
$\gamma$ and $r_{\rm o}$, with their model amplitudes of $-1.89$ and
$5.89$\,Mpc is summarized in Table~\ref{app:params}. In columns $2$--$11$, the
average amplitudes for $\gamma$ and $r_{\rm o}$ of $25$ data sets are shown.
In the last two columns, the rms scatter averaged over all the mask +
filter settings is given. 

\begin{figure}
   \centering
   \includegraphics[width=1.00\linewidth]{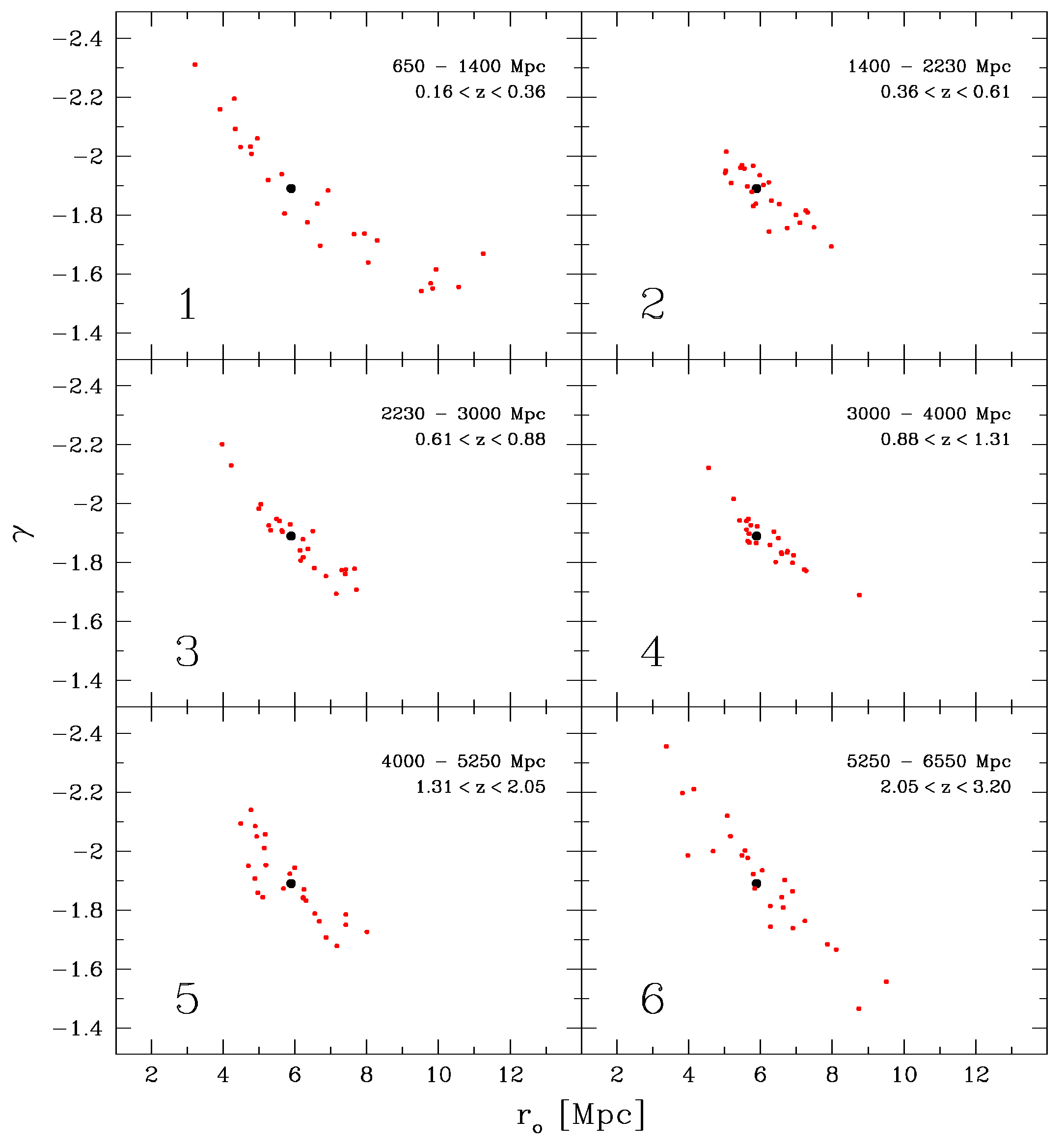}
   \caption{Estimates of the power-law parameters of the correlation
   functions for $25$ sets of simulated data with the `bright star'
   mask and `fluctuation filter' F$_{0.05}$. The input CF parameters
   are indicated by large full dots; parameters extracted using the
   present method are shown with small dots. See text for details.}
   \label{app:rc_gamma_1}
\end{figure}

We note that, in fact, the mask and filters do not introduce any systematic
corrections to our estimates of the CF parameters. One could expect a
moderate raise of uncertainties because both modifications add an additional
noise to the data.  However, the effect is small and the total scatter is
dominated by the stochastic nature of the simulated catalogues. In
Fig.~\ref{app:rc_gamma_1} the distributions of $25$ runs with the BS mask and
the fluctuation filter F$_{0.05}$ are shown in the $r_{\rm o}$--$\gamma$
plane. It appears that the distributions of both parameters  are roughly
centred on their original amplitudes.  Although, some bias towards flatter
slopes and larger correlation length is visible in distance bins that contain
lower number of objects. 

\begin{table*}
\caption{Parameter estimation applied to the simulated, synthetic catalogues of photometric redshifts.}
\label{app:params}
\begin{tabular}{@{}cccccccccccccccccc}
Distance &
\multicolumn{2}{c}{No mask}& &
                 \multicolumn{2}{c}{BS} & &
                   \multicolumn{2}{c}{BS + F$_{0.025}$} & &
                                    \multicolumn{2}{c}{BS + F$_{0.05}$} & &
                                                    \multicolumn{2}{c}{BS + F$_{0.10}$} & &
                                                                                    \multicolumn{2}{c}{rms} \\
  bin    &$\gamma$&$r_{\rm o}$& & 
                   $\gamma$&$r_{\rm o}$ & &
                                   $\gamma$&$r_{\rm o}$ & &
                                                   $\gamma$&$r_{\rm o}$ & &
                                                                   $\gamma$&$r_{\rm o}$ & &$\gamma$&$r_{\rm o}$\\
   1     &  -1.87 & 6.6 & & -1.87 & 6.5 & & -1.89 & 6.3 & & -1.84 & 6.8 & & -1.88 & 6.4 & &   0.22 &   2.2     \\
   2     &  -1.87 & 6.1 & & -1.86 & 6.3 & & -1.87 & 6.1 & & -1.87 & 6.2 & & -1.87 & 6.2 & &   0.10 &   1.1     \\
   3     &  -1.88 & 6.1 & & -1.87 & 6.1 & & -1.88 & 6.1 & & -1.88 & 6.1 & & -1.87 & 6.1 & &   0.11 &   0.9     \\
   4     &  -1.86 & 6.3 & & -1.86 & 6.3 & & -1.87 & 6.3 & & -1.88 & 6.2 & & -1.86 & 6.3 & &   0.09 &   0.9     \\
   5     &  -1.87 & 6.0 & & -1.90 & 5.8 & & -1.88 & 5.9 & & -1.89 & 5.9 & & -1.90 & 5.9 & &   0.15 &   1.1     \\
   6     &  -1.88 & 6.2 & & -1.91 & 6.1 & & -1.90 & 6.1 & & -1.90 & 6.1 & & -1.89 & 6.2 & &   0.21 &   1.6     \\
\end{tabular}

\medskip
{\it Notes.} The simulated data points were drawn according to the underlying probability
distribution that generates the power-law CF with $\gamma = -1.89$
and $r_{\rm o} = 5.9$\,Mpc. BS -- bright star mask, F$_{0.025}$, F$_{0.05}$
and F$_{0.10}$ -- filters; see text for details.
\end{table*}

One should notice that the extent of the scatter plots in
Fig.~\ref{app:rc_gamma_1} results from the statistical character of the
investigated matter. That includes also particular distribution of clusters
in the simulated catalogues. Therefore, the parameter dispersion derived for
the synthetic catalogues is a sum of two components. First, it is generated
by the noise associated with the drawing of points according to the specific
underlying probability distribution. Second, it results from the different
realizations of probability distribution itself. Thus, the scatter plots in
Fig.~\ref{app:rc_gamma_1} do not represent the uncertainties of parameters
estimated from the real data.

A good agreement between the input parameters and their estimates indicates
that the procedures introduced in the paper can be used to derive the space
characteristics of the CF using photometric redshifts. Possible systematic
deviations, if any, are substantially smaller than large statistical
uncertainties.

\label{lastpage}

\end{document}